\documentclass[final,1p,times]{elsarticle}

\usepackage{hyperref}
\usepackage{placeins}
\usepackage{float} 
\usepackage{babel}
\usepackage{babelbib}

\usepackage{microtype}
 \usepackage{multirow} 
\usepackage[table, dvipsnames]{xcolor}
\usepackage{tabularx}
\usepackage{graphicx}
\usepackage{verbatim}
\usepackage{subcaption}
\usepackage{amssymb}
\usepackage{amsmath}
\usepackage{wasysym}
\usepackage{lipsum}
\usepackage{subfiles}
\usepackage{tabu}
\usepackage{gensymb}
\usepackage{array,booktabs} 

\usepackage{lineno}

\journal{Nuclear Physics A}

\begin{document}

\begin{frontmatter}

\title{Polyurethane Scintillators for High-Energy Physics Detectors: Optimization, Scale-up, and Radiation Response}

\author[TECNM]{Rafael Angel Narcio Laveaga}
\author[UAS]{Ildefonso León Monzón}
\author[TECNM]{Juan Carlos Cabanillas Noris}
\author[CTU]{Thanh Luan Nguyen}
\author[CTU,FZU]{Jan Král}
\author[FZU]{Vítězslav Jarý}
\author[CINVESTAV]{Gerardo Herrera Corral}
\author[CERN,CTU]{Vojtech Zabloudil}
\author[CERN]{Etiennette Auffray}          
\author[UAS]{Carlos Duarte Galván} 
\author[CTU]{Solangel Rojas Torres\corref{mycorrespondingauthor}}
\cortext[mycorrespondingauthor]{Corresponding author}

\affiliation[TCNM]{organization={Tecnológico Nacional de México, Instituto Tecnológico de Culiacán},
            addressline={Juan de Dios Bátiz No. 310 Pte. , Col. Guadalupe}, 
            city={Culiacán},
            postcode={80220}, 
            state={Sinaloa},
            country={Mexico}}

\affiliation[UAS]{organization={Facultad de Ciencias Físico Matemáticas, Universidad Autónoma de Sinaloa},
            addressline={Calzada de las Américas y Calle Universitarios, s/n. Ciudad Universitaria}, 
            city={Culiacán},
            postcode={80013}, 
            state={Sinaloa},
            country={Mexico}}

\affiliation[CTU]{organization={Faculty of Nuclear Sciences and Physical Engineering, Czech Technical University},
            addressline={Břehová 7}, 
            city={Prague},
            postcode={115 19}, 
            state={Bohemia},
            country={Czech Republic}}

\affiliation[FZU]{organization={Institute of Physics of the Czech Academy of Sciences},
            addressline={Cukrovarnická 10}, 
            city={Prague},
            postcode={162 00}, 
            state={Bohemia},
            country={Czech Republic}}

\affiliation[CINVESTAV]{organization={Departamento de Física, Centro de Investigación y Estudios Avanzados},
            addressline={Av. Instituto Politécnico Nacional No. 2508, Col. San Pedro Zacatenco}, 
            city={Alcaldía Gustavo A. Madero},
            postcode={07360},
            state={Mexico City},
            country={Mexico}}

\affiliation[CERN]{organization={European Organization for Nuclear Research (CERN)}, addressline={1211 Geneva 23, Switzerland}}

   \begin{abstract}

    Plastic scintillators remain attractive for large-area particle detectors because of their fast response, mechanical flexibility, and relatively simple fabrication. Motivated by future applications including the ALICE~3 Muon Identifier (MID) and the upgrade of the ALICE Fast Interaction Trigger (FIT), this work investigates polyurethane (PU)-based plastic scintillators doped with 2,5-diphenyloxazole (PPO) and 1,4-bis(5-phenyloxazol-2-yl)benzene (POPOP). Systematic information on the optimization and radiation response of this material system remains limited. A series of compositions was therefore fabricated and characterized using minimum-ionizing particles (MIP), optical spectroscopy, and time-correlated single-photon counting. The measurements show that the scintillation response is governed primarily by the PPO concentration, while POPOP mainly modifies the spectral response, and fast effective decay times of approximately 1.6--1.75~ns are maintained across the doped samples. The best-performing formulation was subsequently scaled to 20-$cm$ and 1-m scintillator bars and successfully operated with a single embedded wavelength-shifting fiber and SiPM readout, demonstrating its suitability for meter-scale light collection. Proton-irradiation measurements at the CERN IRRAD facility further revealed systematic degradation of the optical and scintillation responses with increasing exposure, together with substantial recovery after approximately two months. These results demonstrate that optimized PU--PPO--POPOP scintillators can combine fast timing, high light output, scalable detector geometry, and recoverable radiation response, supporting their further development for future large-area particle-detector systems.
    
    \end{abstract}

    \begin{keyword}
    Plastic Scintillator \sep Polyurethane \sep Scintillation Detector \sep Radiation Hardness \sep Minimum-Ionizing Particles \sep High Energy Physics  
    \end{keyword}


\end{frontmatter}

\clearpage
\tableofcontents

\clearpage

\section{Introduction}
\label{introduction}

Plastic scintillators were invented in the early 1950s \cite{schorr_solid_1950} and are a well-established technology in high-energy physics. They are relatively easy to produce and have low maintenance expenses \cite{shevelev_optical_2025}. Some commonly used polymer matrices for making plastic scintillators include polystyrene, polyvinyl toluene, and poly(methyl methacrylate) \cite{hu_preparation_2023}, and other materials such as epoxy-resin have also been tested \cite{nam_study_2017}. Recently, due to the increase in luminosity of High Energy Physic (HEP) experiments, the need to generate materials more resistant to radiation has arisen, and  other polymer bases are being explored. 

Polyurethane (PU) has stood out as a robust alternative resistant to environmental conditions \cite{shevelev_optical_2025,shevelev_new_2023,syntfeld-kazuch_characterization_2024,maiatska_polyurethane-based_2026}. This material has also been tested as candidate for being used as a host matrix for nanocrystals and quantum dots \cite{kral_tailored_2026}.

Typically, the polymer matrices on their own have low light yield, making adding additional dyes, like 2,5-Diphenyloxazole (PPO), a requirement to improve their scintillation performance. Due to the emission peak resulting from adding a primary dye usually being in the ultraviolet (UV), to reduce self-absorption and for better spectral matching with photodetectors, a secondary dye, like 1,4-bis(5-phenyloxazol-2-yl) benzene (POPOP), can also be added \cite{shevelev_optical_2025,younesi_effect_2024,alex_effect_2022}, with the aforementioned two dyes being some of the most common dopants used for making plastic scintillators \cite{moore_high-resolution_2025} and those are the dyes selected for this work.


The objective of this work is to develop and characterize a new PU-based scintillator material suitable for use in new radiation detectors in HEP and capable of competing with what has been reported in the literature while maintaining a relatively low dopant concentration compared to typical formulations reported for PU-based scintillators (up to $20-30~wt\%$ of primary dye) \cite{shevelev_optical_2025,maiatska_polyurethane-based_2026}. Since high dopant concentrations have been reported to make the material more prone to aging \cite{younesi_effect_2024} and more susceptible to radiation damage \cite{MIZUEHAMADA1999148}, lower dopant concentrations were selected to enhance radiation hardness. Part of the motivation for this work is to obtain a material that would be a suitable alternative for future detectors under development, like the MID for ALICE 3, and the upgrade of the FIT detectors in the ALICE experiment. 
Nevertheless, other experiments around the world use plastic scintillator detectors and could benefit from alternatives for their design and construction. A few examples include the CMS hadron end-cap calorimeter~\cite{Sirunyan_2020}, the STAR event plane detector~\cite{ADAMS2020163970}, the Belle-II scintillator end caps~\cite{Belle-II:2010dht}, and the T2K INGRID detector~\cite{T2K:2011qtm}. This trend is expected to continue in the future with, for example, the BeBe detector at NICA~\cite{ALVARADO2020163150} and the muon detector for the SHiP experiment at CERN~\cite{Baldini_2017}.

In this work, we studied the effects of low concentrations of dopants in PU, first finding an optimal concentration in terms of its relative light output compared to the sample with the highest light yield, then measuring its transmission and emission spectra, its decay time, and preparing a scintillator bar with the optimal concentration to compare its light output to other scintillators reported in the literature. Additionally, a radiation test was performed on the plastic scintillator used to make the bars and the pure host matrix to see the effects of proton radiation on the material.

\section{Sample Fabrication}
\label{sec:Materials}

\begin{figure}[H]
  \centering
    \includegraphics[width=0.38\linewidth]{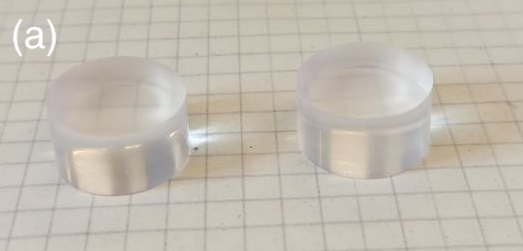} \includegraphics[width=0.35\linewidth]{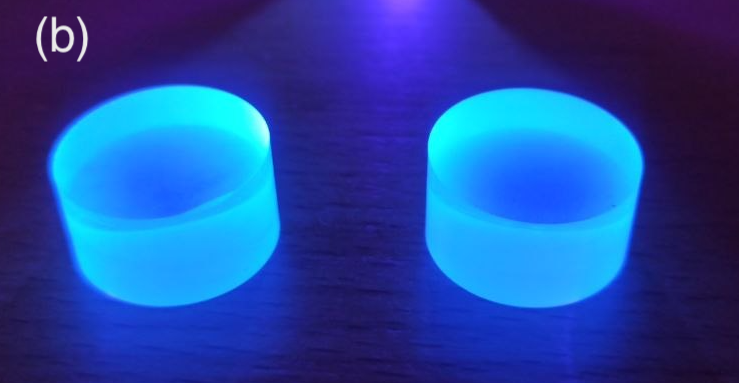}
    \caption{Pictures of two samples under (a) ambient and (b) UV light. The sizes are of $\phi~21\times10~mm^3$.}
    \label{fig:PU_sample}
\end{figure}

The UV-resistant aliphatic polyurethane polymer matrix was prepared by resin casting from the curative and pre-polymer components (Techno Cast 3284, Polymères Technologies) at a ratio of 1:1 at room temperature \cite{polymeres_technologies_100_nodate}. The PPO and POPOP dopants were added as primary and secondary, respectively. The PPO dopant was mixed and stirred into the curative component. The mixture was subsequently heated at 80~$^\circ$C for 16~hours. Afterward, POPOP was integrated into the blend and stirred. It was then heated again at 60~$^\circ$C for an additional 8~hours and stirred once more to ensure homogeneity. To initiate the polymerization process, the mixed curative component was added to the pre-component and stirred for 2~minutes. It was placed in a vacuum chamber to eliminate air bubbles and cast into a cylindrical mold, which was then maintained inside a pressure chamber for 24-48~hours at 50~psi ($\sim$3.45 bar). After polymerization, the plastic scintillator was extracted from the mold, then cut and polished it to approximately $\phi~21\times10~mm^3$, as shown in Fig.~\ref{fig:PU_sample}.

The PPO concentrations ranged from $0.4~wt\%$ to $5~wt\%$, and POPOP concentrations ranged from $0.016~wt\%$ to $0.07~wt\%$. A pure polyurethane sample (undoped) was made as a reference. The samples were labeled according to their dopant percentages: P\textit{x}S\textit{y}, where \textit{x} and \textit{y} are the percentage numbers (following the decimal point) of the primary and secondary dopants, respectively. The full nomenclature of the samples is detailed in Table~\ref{tab:Samples}. The samples were prepared and tested batch by batch to optimize the dopant concentrations. Based on the results of each batch, the PPO concentration was progressively increased, while POPOP was maintained at approximately $1~wt\%$ relative to PPO.

\begin{table}[htbp]
\centering
\small
\setlength{\tabcolsep}{6pt}
\renewcommand{\arraystretch}{1.15}
\caption{Compositions and labels of the prepared samples. The POPOP/PPO ratio is calculated from the corresponding weight concentrations.}
\label{tab:Samples}
\begin{tabularx}{\linewidth}{*{4}{>{\centering\arraybackslash}X}}
\hline
\textbf{PPO} & \textbf{POPOP} & \textbf{POPOP/PPO} & \textbf{Sample ID}\\
\textbf{(wt\%)} & \textbf{(wt\%)} & \textbf{(\%)} & \\
\hline
0.0 & 0.000 & --   & P0S0 \\
\hline
0.4 & 0.020 & 5.00 & P04S020 \\
\hline
\multirow{2}{*}{0.8} & 0.016 & 2.00 & P08S016 \\ \cline{2-4}
                   & 0.020 & 2.50 & P08S020 \\
\hline
\multirow{2}{*}{1.2} & 0.016 & 1.33 & P12S016 \\ \cline{2-4}
                   & 0.020 & 1.67 & P12S020 \\
\hline
\multirow{2}{*}{1.6} & 0.016 & 1.00 & P16S016 \\ \cline{2-4}
                   & 0.020 & 1.25 & P16S020 \\
\hline
\multirow{2}{*}{2.0} & 0.016 & 0.80 & P20S016 \\ \cline{2-4}
                   & 0.020 & 1.00 & P20S020 \\
\hline
\multirow{5}{*}{3.0} & 0.016 & 0.53 & P30S016 \\ \cline{2-4}
                   & 0.024 & 0.80 & P30S024 \\ \cline{2-4}
                   & 0.030 & 1.00 & P30S030 \\ \cline{2-4}
                   & 0.036 & 1.20 & P30S036 \\ \cline{2-4}
                   & 0.042 & 1.40 & P30S042 \\
\hline
\multirow{5}{*}{4.0} & 0.024 & 0.60 & P40S024 \\ \cline{2-4}
                   & 0.032 & 0.80 & P40S032 \\ \cline{2-4}
                   & 0.040 & 1.00 & P40S040 \\ \cline{2-4}
                   & 0.048 & 1.20 & P40S048 \\ \cline{2-4}
                   & 0.056 & 1.40 & P40S056 \\
\hline
\multirow{5}{*}{5.0} & 0.030 & 0.60 & P50S030 \\ \cline{2-4}
                   & 0.040 & 0.80 & P50S040 \\ \cline{2-4}
                   & 0.050 & 1.00 & P50S050 \\ \cline{2-4}
                   & 0.060 & 1.20 & P50S060 \\ \cline{2-4}
                   & 0.070 & 1.40 & P50S070 \\
\hline
\end{tabularx}
\end{table}

\begin{figure}[H]
  \centering
  \begin{subfigure}[t]{0.9\textwidth}
    \centering
    \includegraphics[width=\linewidth]{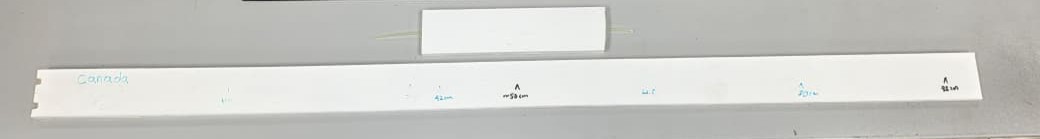}
    \caption{Top view of the 20~$cm$ and 1~m PU bars}
    \label{fig:PU_bar_Top}
  \end{subfigure}\hfill
  \begin{subfigure}[t]{0.48\textwidth}
    \centering
    \includegraphics[width=\linewidth]{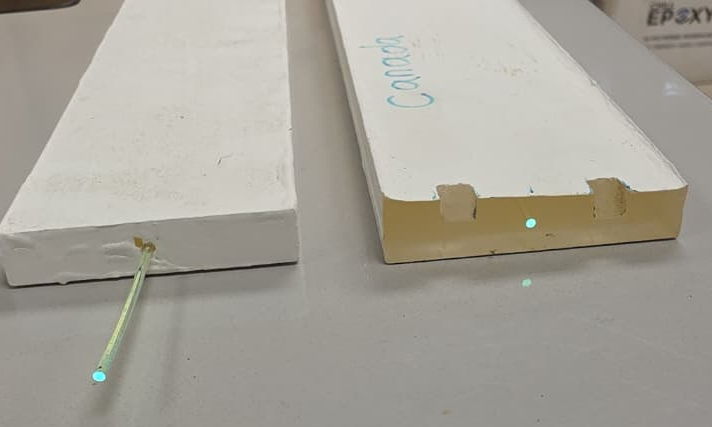}
    \caption{Front view of the 20~$cm$ and 1~m PU bars}
    \label{fig:PU_bars_Front}
  \end{subfigure}
  \caption{Pictures of the scintillator bars of (a) $200~mm$ and (b) $1000~mm$ length made with PU and vcovered with $TiO_2$ reflector. Both have a width of $50~mm$ and a thickness of $10~mm$. }
  \label{fig:PU_bars_pics}
\end{figure}

Two scintillator bars were fabricated using the same technique described earlier, based on the composition of the best performing sample. One of $200\times 50 \times 10 ~mm^3$ and a second one with $1000\times 50 \times 10 ~mm^3$.  Each bar was constructed with a Kuraray Y11 (200) wavelength-shifter (WLS) with a $\phi~1.5$ diameter fiber centered along the $ 20~mm$ and $1000~mm$ lengths of the corresponding bar. The embedding of the WLS fibre is different for each bar: the 200~$mm$ bar was produced with a 2~$mm$ diameter cavity to insert the fibre (therefore is a thin air layer interface between the bar and the fibre), and the 1000~$mm$ bar was produced with the fibre embedded during the polymerization. The surfaces were polished and prepared for coating with $TiO_{2}$ layers as a reflector. The $TiO_{2}$ covers all surfaces of the bar except for one end of the WLS fiber, which was used to couple to the photodetector. Fig.~\ref{fig:PU_bars_pics} shows both PU bars. 

\section{Experimental Measurements}
\label{sec:Experiments}

\subsection{Transmittnace and Emission Measurements}

The transmittance and photoluminescence emission of the samples were measured using a FluoroMax 4-Plus spectrofluorometer (Horiba Jobin Yvon). The transmittance was measured using the corresponding accessory by comparing the light intensity between the sample and the blank. This measurement was performed from 200 nm to 800 nm with a step of 1 nm. The emission spectra were measured based on the default setup. The excitation wavelength ($\lambda_{exc}$) at $280~nm$ was selected for the emission measurement of doped samples. For the case of the undoped sample, the excitation wavelength of $270~nm$ was used.

The transmission edge was characterized from the wavelength derivative of the measured transmittance spectrum. The edge wavelength, $\lambda_{\rm edge}$, was defined as the wavelength corresponding to the maximum positive slope,

\begin{equation}
\lambda_{\rm edge}
=
\operatorname*{arg\,max}_{\lambda}
\left[
\frac{dT(\lambda)}{d\lambda}
\right],
\end{equation}
where $T(\lambda)$ is the measured transmittance. The spectral width of the transmission edge, $\Delta\lambda_{\rm edge}$, was determined as the full width at half maximum (FWHM) of the corresponding derivative peak.

The emission maximum, $\lambda_{\rm em}$, was determined individually from the measured emission spectrum of each sample. The transmittance at the emission maximum, $T(\lambda_{\rm em})$, was obtained by interpolation of the measured transmittance spectrum at $\lambda_{\rm em}$. To account for the finite width of the emission band, an emission-weighted effective transmittance was calculated as

\begin{equation}
T_{\rm eff}
=
\frac{
\displaystyle
\int_{\lambda_1}^{\lambda_2}
T(\lambda) I_{\rm em}(\lambda)\,d\lambda
}{
\displaystyle
\int_{\lambda_1}^{\lambda_2}
I_{\rm em}(\lambda)\,d\lambda
},
\end{equation}
where $I_{\rm em}(\lambda)$ is the measured emission intensity.
The integration limits $\lambda_1$ and $\lambda_2$ define the spectral
interval retained for the analysis within the common wavelength range
of the transmittance and emission measurements.

\subsection{Time-Resolved Photoluminescence Measurements}

The time-resolved photoluminescence measurements were performed using the time-correlated single photon counting (TCSPC) method  with the custom-made 5000M spectrofluorometer (Horiba) \cite{brosi_time-correlated_2025}. A nanosecond nanoLED pulsed light source (Horiba) was used for excitation. A single grating monochromator and a TBX-04 Horiba PMT were used for detection. The excitation and monitor wavelengths ($\lambda_{mon}$) were set based on the photoluminescence emission spectrum. The doped samples were excited with a $\lambda_{exc}=~389~nm$ at $80~kHz$ with $\lambda_{mon}=425~nm$, while the blank sample was excited with a $\lambda_{exc}=281~nm$ at $500~kHz$ with $\lambda_{mon}=310~nm$.

The measured TCSPC spectrum was described by a multi-exponential decay
function convolved with the instrument response function (IRF), represented
by a Gaussian function, together with a constant background, as follows:
\begin{equation}
I(t)=
G(t;t_0,\sigma_{\rm IRF})
\otimes
\left[
\sum_i A_i
\exp\left(-\frac{t}{\tau_i}\right)
\right]
+C,
\label{eq:conv_decay}
\end{equation}
where $t$ is the time; $t_0$ and $\sigma_{\rm IRF}$ are the position and
width of the Gaussian IRF, respectively; $A_i$ and $\tau_i$ are the
amplitude and decay time of the $i$th decay component; $C$ is the constant
background; and $\otimes$ denotes convolution. The Gaussian IRF parameters
were determined from the measured IRF for each experimental setting and
used in the corresponding decay fit. The fractional contribution of each
decay component was calculated as
\begin{equation}
R_i=
\frac{A_i\tau_i}
{\sum_j A_j\tau_j},
\label{eq:Ri}
\end{equation}
where $R_i$ represents the contribution of the $i$th decay component.
The one- and three-component fittings were applied for the case of doped and undoped samples, respectively \cite{shevelev_new_2023}. The effective decay time is determined by: 
\begin{align}
    \tau_{eff} &= \left(\sum_{i} \frac{R_{i}}{\tau_{i}} \right)^{-1}
    \label{eq:eff_tau_weight}
\end{align}

\subsection{MIP-based Light Output Measurements}

\begin{figure}[h]
    \centering
    \includegraphics[width=0.65\linewidth]{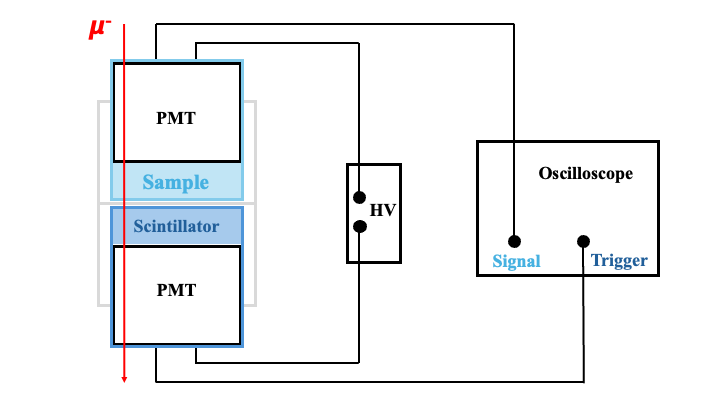}
    \caption{Cylindrical couple used for the measurement of the relative light output of the samples}
    \label{fig:Samples_Workbench}
\end{figure}

The MIP-based light output was measured using cosmic rays, as illustrated in Fig.~\ref{fig:Samples_Workbench}. The plastic scintillators were coupled to the Photonis VS202K/02 PMT. The CAEN V6533M module was used for high voltage supply, which was controlled via the CAEN V1718 module. The sample and trigger signals were connected to an HD04104-MS Teledyne Lecroy oscilloscope. The triggered signals were collected for approximately 24 hours. The collected raw signals were analyzed to obtain the charge. A dynamic integration window was defined around the maximum amplitude, with 25 ns pre-window and 50 ns post-window. 

The charge spectrum was described by four components, including a Gaussian pedestal, a Gaussian PMT-noise contribution, a Landau--Gaussian convolution describing the MIP response, and an exponential background as follows:
\begin{align}
    F(Q)= N_{\rm ped}G(Q;\mu_{\rm ped},\sigma_{\rm ped}) 
    +N_{\rm noise}G(Q;\mu_{\rm noise},\sigma_{\rm noise})
    +N_{\rm MIP}(L\!\otimes\!G)(Q;Q_{\rm MIP})
    +N_{\rm bg}e^{-\lambda Q}
    \label{eq:MIP_q}
\end{align}
In which, $Q$ is the integrated charge; $\mu_{\rm ped}$ and $\mu_{\rm noise}$ are the mean positions of the pedestal and noise components, respectively; $\sigma_{\rm ped}$ and $\sigma_{\rm noise}$ are their Gaussian widths; $Q_{\rm MIP}$ represents the most probable MIP charge; $\lambda$ is the exponential background slope; and $N_i$ denotes the normalization of each component. The most probable value (MPV) was expected from the fit. The absolute MIP position was then determined by subtracting the pedestal mean ($\mu_{ped}$) from the extracted MPV. It should be noted that all charge measurements were normalized to the thickness ($L$) of each sample.

\begin{figure}[h]
    \centering
    \includegraphics[width=0.8\linewidth]{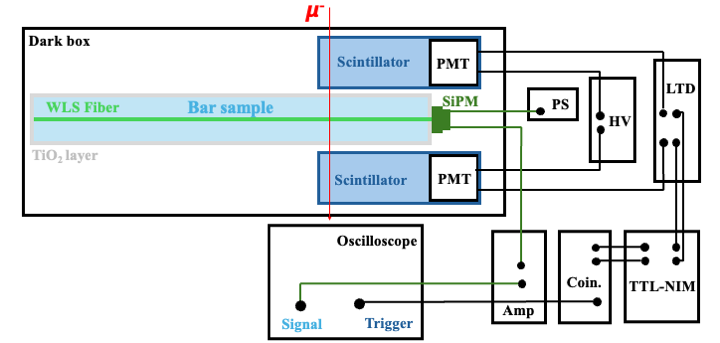}
    \caption{Transversal view of the workbench for the scintillator bars.}
    \label{fig:Bars_Workbench}
\end{figure}

The fabricated bars were measured relative to the MIP position using the setup shown in Fig.~\ref{fig:Bars_Workbench}. The WLS fiber output of the scintillator bar was coupled to a Hamamatsu S13360-3075PE SiPM with a thin layer of optical grease to improve light collection. 
Two trigger detectors were set up, each consisting of a scintillator attached to a PMT, for MIP-based light-output measurement, powered by the same high-voltage module. 
The signals from the trigger detectors were connected to the  CAEN V814 Low Threshold Discriminator (LTD) module, then to a custom TTL to NIM converter to send the signal to the CAEN V976 coincidence module. The signal from the SiPM was amplified 10 times by the CAEN V975 8 Channel Fast Amplifier. Both the SiPM and trigger coincidence signals were connected to the HD04104-MS Teledyne LeCroy oscilloscope.

The coincidence signals were collected and used to construct the charge spectrum. The charge integration window was 230 ns with a pre-window of 30 ns. The $Q_{\rm MIP}$ and $\sigma_{\rm MIP}$ were determined from charge fitting. The MIP position was measured in terms of the detected photoelectrons (phe), which were normalized to the bar thickness. A single-photon counting calibration was performed. $Q_{\rm MIP}$ was divided by the gain to calculate the amount of photoelectrons detected.

\subsection{Radiation Hardness Measurements}
\label{ssec:RadExp}
\begin{figure}[h]
  \centering
  \includegraphics[width=0.45\textwidth]{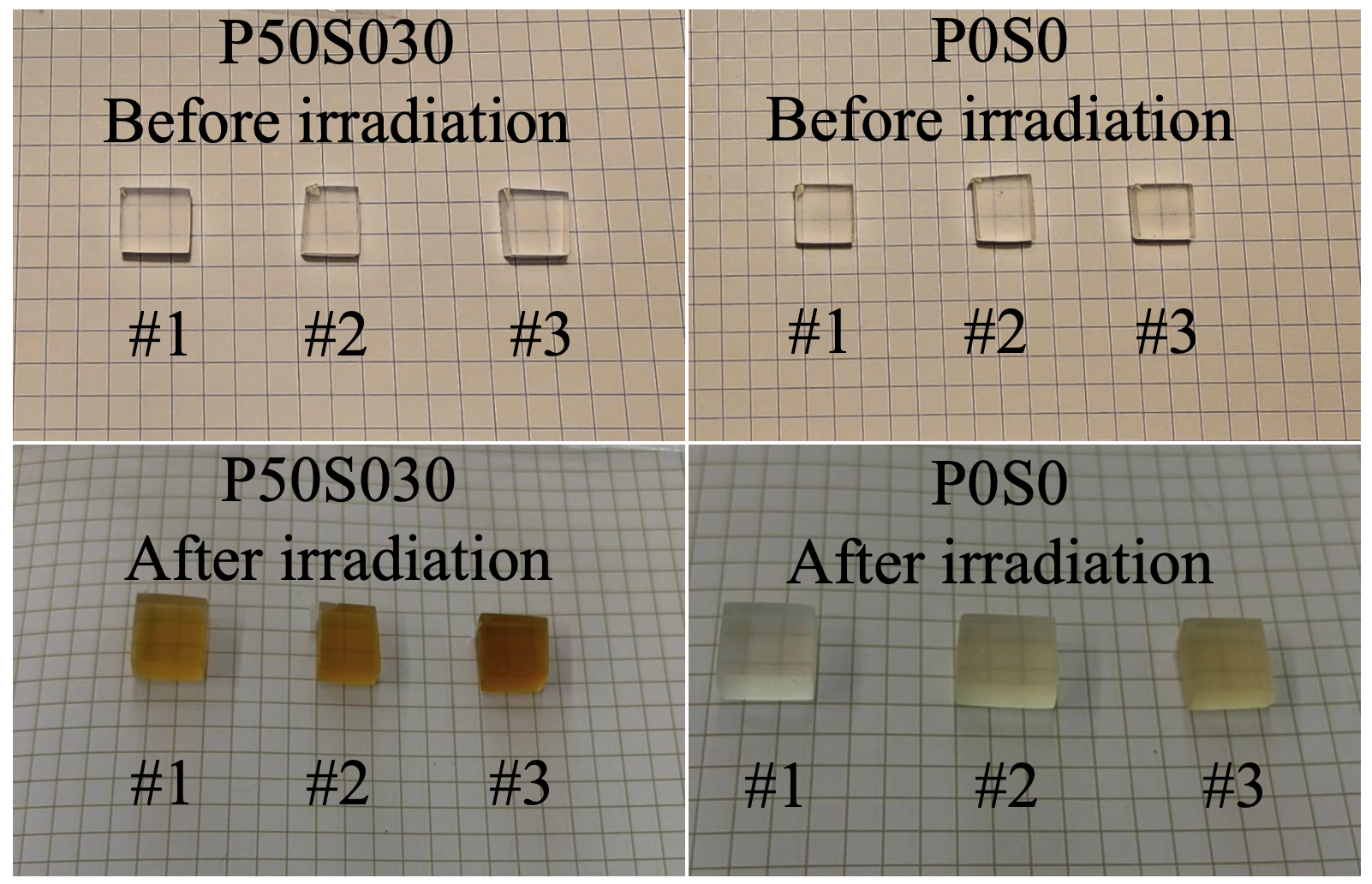}
  \caption{Samples before and after proton irradiation.}
  \label{fig:sample_beforeafter}
\end{figure}

To investigate the radiation hardness of PU-based materials in a high-radiation environment, the P0S0 and P50S030 samples were prepared for proton irradiation at PS-IRRAD (CERN) \cite{Gkotse:2896062}. Transmittance, photoluminescence, and light yield were measured before and after irradiation at CERN. The requested fluences were $1.99\times 10^{13}$, $9.99\times 10^{13}$, and $1.99\times 10^{14}\,\mathrm{protons\,cm^{-2}}$, corresponding to expected doses of 1~Mrad, 5~Mrad, and 10~Mrad, or low, intermediate, and high levels of proton exposure, respectively. Each batch was cut into 4 samples with a thickness of 5~$mm$, and the sample surfaces covered the proton-beam cross section. The P50S030 batch was used to evaluate changes in optical transmittance, photoluminescence, and light yield before and after irradiation, while for the P0S0 batch, only optical transmittance was measured. One sample was retained as an unirradiated pristine reference, while the other three were subjected to different irradiation levels. The samples were numbered 1, 2, and 3, corresponding to the expected doses for the low, intermediate, and high levels of proton exposure and their properties where measured before, just after irradiation and after two months of irradiation. Photographs of the samples before and after irradiation are shown in Fig.~\ref{fig:sample_beforeafter}.

Transmittance was measured using a PerkinElmer Lambda 650 UV-Vis spectrometer in double-beam mode with an integrating sphere. Note that the integrating sphere collects not only the directly transmitted beam but also part of the diffusely scattered light; therefore, the measured transmittance may differ from that obtained using a conventional setup (e.g., a Fluoromax R928P spectrofluorometer). To quantify the effect of irradiation, the radiation-induced absorption coefficient was calculated from the transmittance spectra measured before and after irradiation as:
\begin{equation}
    \mu_{\mathrm{ind}}(\lambda) = \frac{1}{d} \ln\!\left( \frac{T_0(\lambda)}{T(\lambda)} \right),
    \label{eq:induced_absorption}
\end{equation}
where $d$ is the sample thickness, and $T_0(\lambda)$ and $T(\lambda)$ are the transmittances before and after irradiation, respectively.

Photoluminescence was measured using a PerkinElmer LS-55 luminescence spectrometer. The sample holder was designed to fix the sample position and orientation and to ensure excitation and emission from the central region of the sample. Light yield was evaluated from the Compton edge using a $^{137}$Cs source. Bare samples were placed at the center of the window of a Hamamatsu H3177-51 PMT. The scintillation signal was collected, and charge spectra were constructed using a CAEN DT5720A digitizer with a 200~ns integration window.

\section{Results and Discussion}
\label{sec:Results}

\subsection{Transmittance and photoluminescence emission spectra}
\label{subsec:Optical_Res}

\begin{figure}[h]
    \centering
    \includegraphics[width=0.9\linewidth]{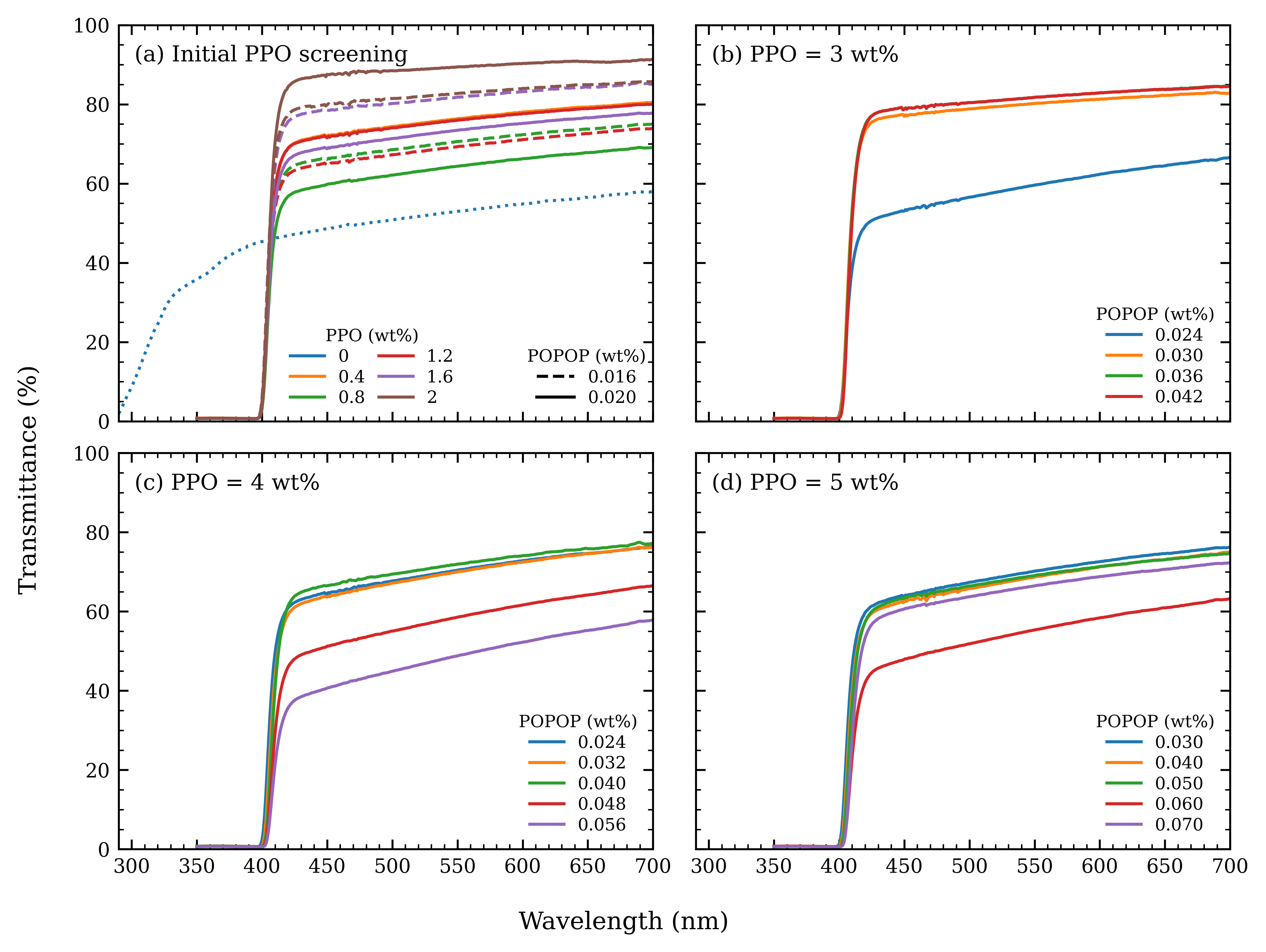}
    \caption{
    Transmittance spectra of plastic scintillator samples with varying PPO and POPOP concentrations. (a) Initial PPO screening (0 to 2~$wt\%$) at two POPOP concentrations (0.016 and 0.020~$wt\%$, dashed and solid lines, respectively). (b)–(d) POPOP concentration scans at fixed PPO loadings of 3, 4, and 5~$wt\%$, respectively. POPOP concentrations for each panel are indicated in the legends.
    }
    \label{fig:Trans_doped}
\end{figure}

The transmittance spectra of the investigated samples are shown in Fig.~\ref{fig:Trans_doped}, and the derived optical parameters are summarized in Table~\ref{tab:SampleProperties}. For the doped samples, $\lambda_{\rm edge}$ ranges from 404 to 410~nm, while $\Delta\lambda_{\rm edge}$ remains within 13.3--14.2~nm. Comparison with the compositions listed in Table~\ref{tab:Samples} shows that the transmission edge generally shifts toward longer wavelengths with increasing fluor loading, whereas its width changes only moderately. In contrast, the undoped P0S0 sample exhibits a shorter-wavelength and
broader transmission edge, with $\lambda_{\rm edge}=309$~nm and $\Delta\lambda_{\rm edge}=40.5$~nm.

\begin{figure}
\centering
\includegraphics[width=0.9\linewidth]{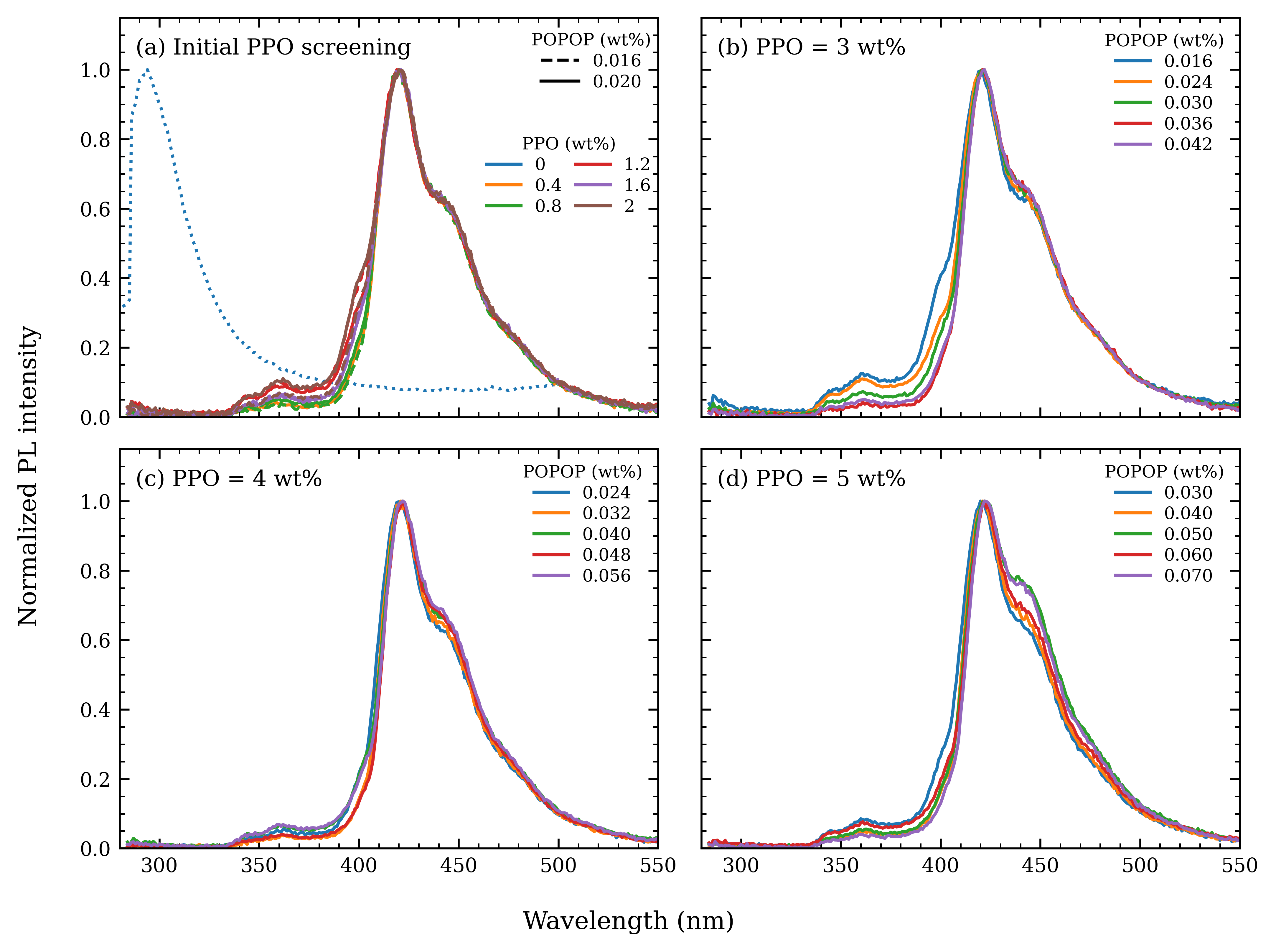}
\caption{
    Emission spectra measured with excitation wavelengths of $280$ nm and $270$ nm of the plastic scintillator samples with varying PPO and POPOP concentrations. (a) Initial PPO screening (0 to 2~$wt\%$) at two POPOP concentrations (0.016 and 0.020~$wt\%$, dashed and solid lines, respectively). (b)–(d) POPOP concentration scans at fixed PPO loadings of 3, 4, and 5~$wt\%$, respectively. POPOP concentrations for each panel are indicated in the legends.
}
\label{fig:Emission280nm}
\end{figure}

The corresponding emission spectra are shown in Fig.~\ref{fig:Emission280nm}. P0S0 exhibits an intrinsic emission maximum at approximately 294~nm, whereas the doped samples are dominated by POPOP-related emission at approximately 419--423~nm. A weaker PPO-related feature is consistently observed at approximately 358--363~nm, indicating that residual PPO fluorescence remains present in the doped samples. The relative contribution of this feature varies with the PPO to POPOP composition, although its dependence on POPOP concentration is not strictly monotonic. Across the investigated compositions, both the POPOP-dominated emission maximum and the transmission edge show modest shifts toward longer wavelengths. At higher fluor loadings, the transmission onset therefore approaches the short-wavelength side of the POPOP emission band, consistent with increased spectral overlap between the absorption and emission regions.

To account for the finite width of the emission band, the optical transmission was further characterized using the emission-weighted effective transmittance, $T_{\rm eff}$, in addition to $T(\lambda_{\rm em})$. The corresponding values are reported in Table~\ref{tab:SampleProperties}. For most compositions, $T_{\rm eff}$ is lower than $T(\lambda_{\rm em})$, reflecting the contribution of photons on the short-wavelength side of the emission band, where the transmittance decreases toward the transmission edge. Neither $T_{\rm eff}$ nor $T(\lambda_{\rm em})$ varies monotonically with fluor concentration, indicating that the optical transmission of the emitted light depends on the combined evolution of the emission spectrum and the wavelength-dependent transmittance.

\subsection{Time-Resolved Photoluminescence}
\label{subsec:Decay_Time_Res}

\begin{figure}[h!]
\centering
\begin{subfigure}[t]{0.48\linewidth}
  \centering
  \includegraphics[width=\linewidth]{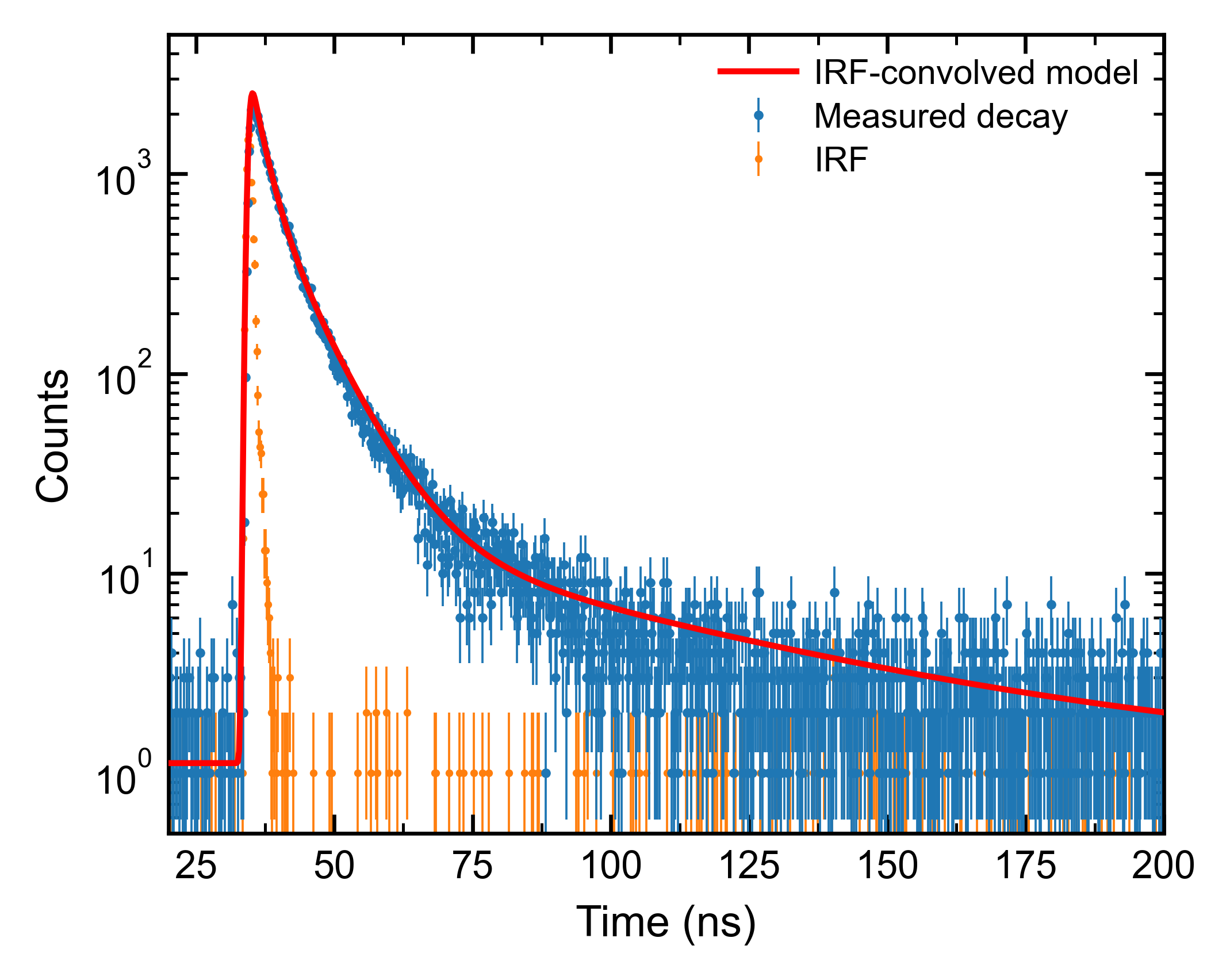}
  \caption{Undoped sample P0S0}
  \label{fig:P0S0_decay}
\end{subfigure}\hfill
\begin{subfigure}[t]{0.48\linewidth}
  \centering
  \includegraphics[width=\linewidth]{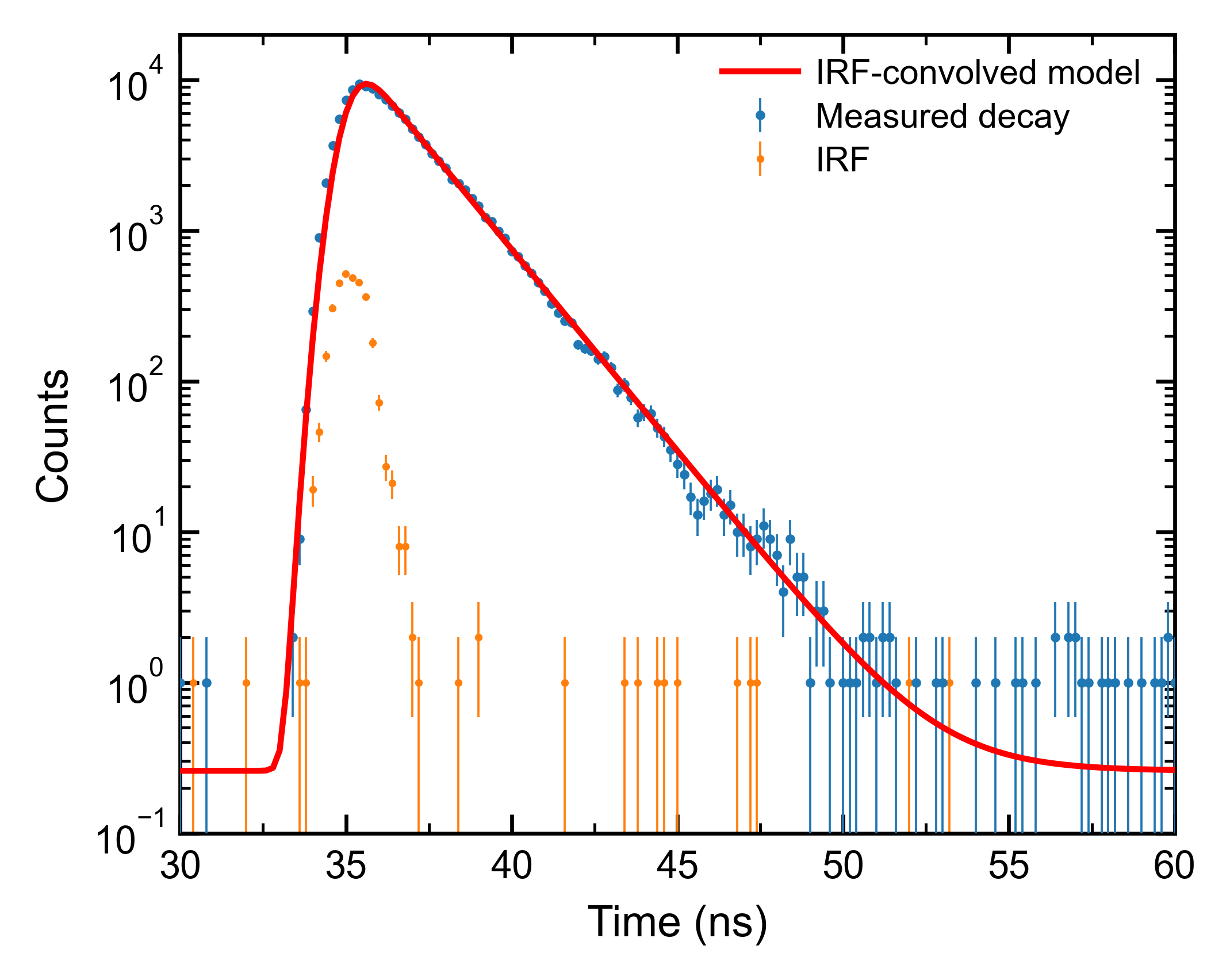}
  \caption{Doped sample P50S030}
  \label{fig:P50S030_decay}
\end{subfigure}
\caption{Time-resolved photoluminescence decays of (a) undoped P0S0 and (b) doped P50S030. Measured data is in blue points, the instrument response in orange points, and IRF-convolved model fit in red line. }
\label{fig:Decay_examples}
\end{figure}

\begin{figure}[h!]
\centering
\begin{subfigure}[t]{0.48\linewidth}
  \centering
  \includegraphics[width=\linewidth]{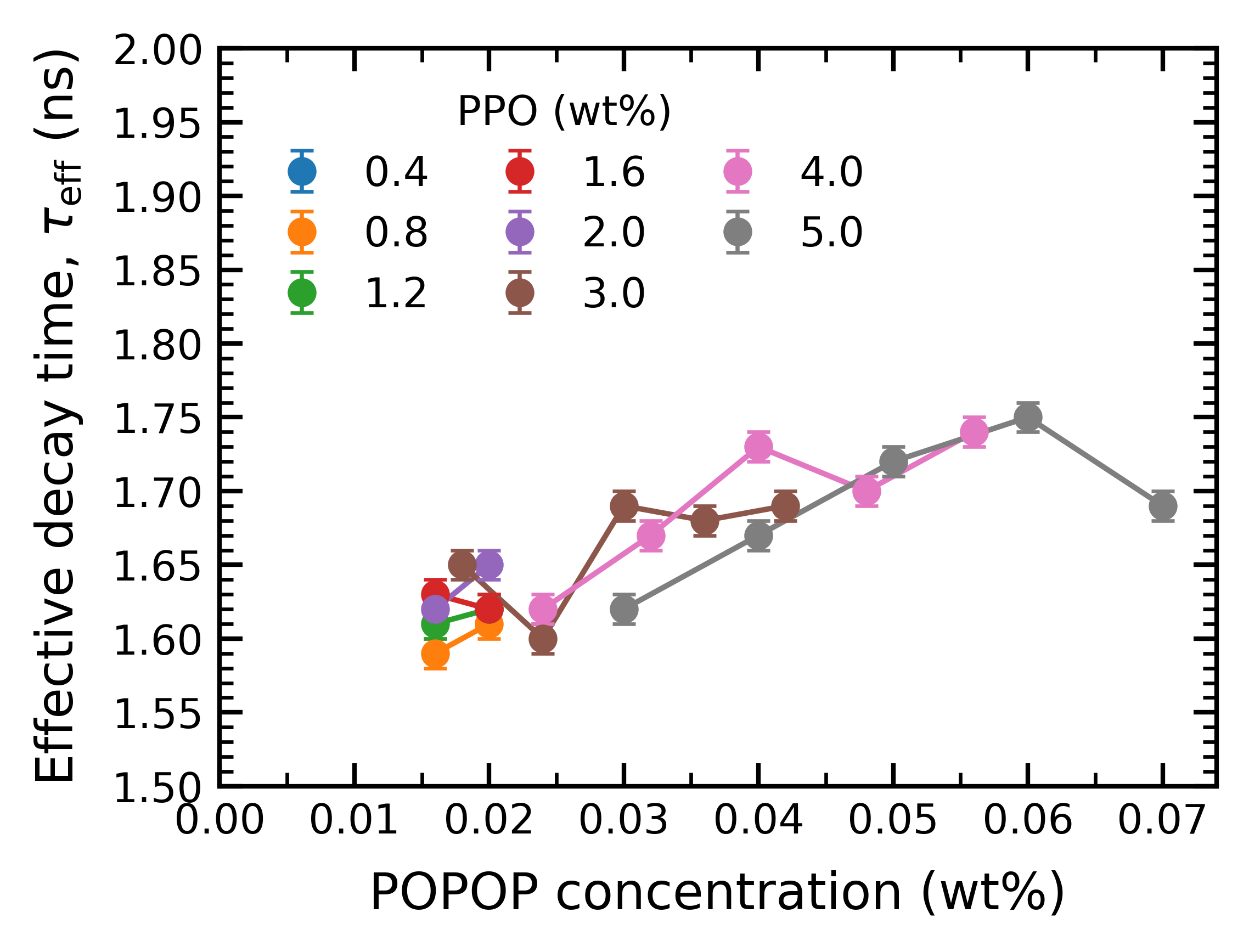}
  \caption{Effective decay time $\tau_{\mathrm{eff}}$ with respect to POPOP concentration}
  \label{fig:tau_eff_POPOP}
\end{subfigure}\hfill
\begin{subfigure}[t]{0.48\linewidth}
  \centering
  \includegraphics[width=\linewidth]{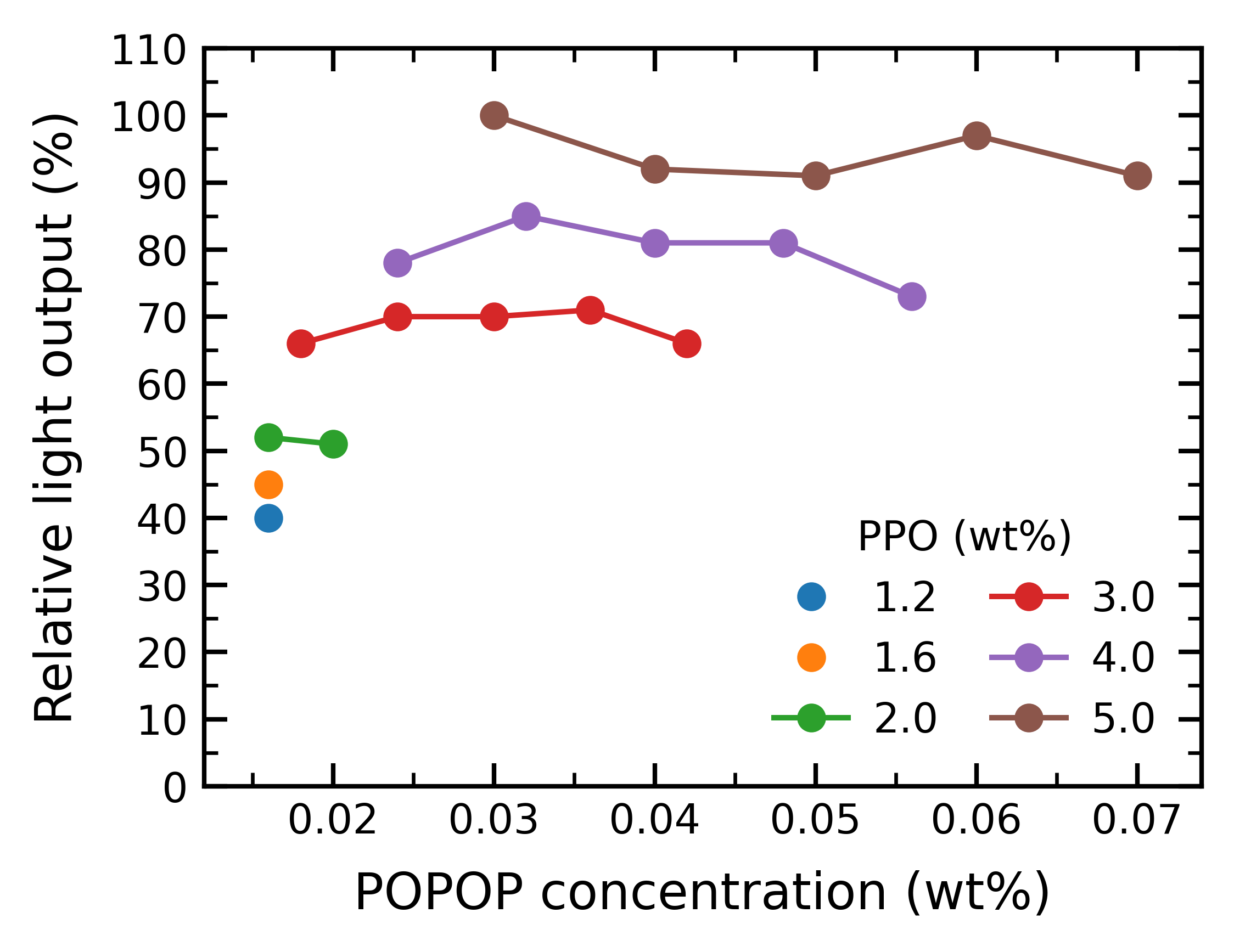}
  \caption{Relative light output $Q_{\mathrm{MIP}}$ with respect to POPOP concentration}
  \label{fig:MIP_q_POPOP}
\end{subfigure}
\caption{Dependence on POPOP concentration: decay-time trend (a) and MIP charge trend (b) for different PPO concentrations.}
\label{fig:tau_eff_and_MIP_POPOP}
\end{figure}

The time-resolved photoluminescence (TRPL) measurements were used to characterize the temporal response of the investigated samples. Representative decay curves and fits are shown in Fig.~\ref{fig:Decay_examples}, while the extracted effective decay times, $\tau_{\rm eff}$, are summarized in Table~\ref{tab:SampleProperties} and plotted in Fig.~\ref{fig:tau_eff_and_MIP_POPOP} 

The undoped P0S0 sample exhibits an effective decay time of $4.10\pm0.01$~ns, markedly longer than those of the PPO--POPOP-doped samples. For the doped samples, $\tau_{\rm eff}$ remains within a comparatively narrow range of $1.59$--$1.75$~ns despite the broad range of PPO and POPOP concentrations investigated (from Table~\ref{tab:Samples}). The introduction of the PPO--POPOP fluor system therefore produces a substantially faster photoluminescence response than that observed for the undoped PU matrix.

Within the doped sample set, the dependence of $\tau_{\rm eff}$ on composition is comparatively weak and is not strictly monotonic. Nevertheless, a modest increase in $\tau_{\rm eff}$ is observed for several of the higher-loading compositions. This behavior is particularly visible within the P50 series, where $\tau_{\rm eff}$ increases from 1.62~ns for P50S030 to 1.75~ns for P50S060 before decreasing to 1.69~ns for P50S070. Similar, although smaller, variations are observed within the P30 and P40 series.

The relatively narrow distribution of $\tau_{\rm eff}$ among the doped samples indicates that changes in fluor composition have a much smaller effect on the temporal response than the initial introduction of the PPO--POPOP fluor system into the PU matrix. At the same time, the non-monotonic variation within the fixed-PPO series suggests that the decay response cannot be described solely by the concentration of either PPO or POPOP. A more detailed interpretation of the composition-dependent decay behavior is considered in the Discussion (Section~\ref{sec:discussion}).


\begin{table}[htbp]
\centering
\small
\setlength{\tabcolsep}{6pt}
\renewcommand{\arraystretch}{1.15}

\caption{Optical properties, effective decay time, and relative light output of the investigated samples: transmission-edge wavelength ($\lambda_{\rm edge}$), transmittance at $\lambda_{\rm edge}$ ($T(\lambda_{\rm edge})$), transmission-edge width ($\Delta\lambda_{\rm edge}$), emission maximum ($\lambda_{\rm em}$), transmittance at $\lambda_{\rm em}$ ($T(\lambda_{\rm em})$), emission-weighted effective transmittance ($T_{\rm eff}$), effective decay time ($\tau_{\rm eff}$), and MIP light output relative to P50S030 (Relative LO).}

\label{tab:SampleProperties}

\resizebox{\linewidth}{!}{%
\begin{tabular}{
*{1}{>{\centering\arraybackslash}p{1.2cm}}
*{3}{>{\centering\arraybackslash}p{1.5cm}}
*{3}{>{\centering\arraybackslash}p{1.6cm}}
*{1}{>{\centering\arraybackslash}p{1.6cm}}
*{1}{>{\centering\arraybackslash}p{1.5cm}}
}
\hline
\textbf{Sample} &
$\boldsymbol{\lambda_{\rm edge}}$ &
$\boldsymbol{T(\lambda_{\rm edge})}$ &
$\boldsymbol{\Delta\lambda_{\rm edge}}$ &
$\boldsymbol{\lambda_{\rm em}}$ &
$\boldsymbol{T(\lambda_{\rm em})}$ &
$\boldsymbol{T_{\rm eff}}$ &
$\boldsymbol{\tau_{\rm eff}}$ &
\textbf{Relative LO}
\\
&
\textbf{(nm)} &
\textbf{(\%)} &
\textbf{(nm)} &
\textbf{(nm)} &
\textbf{(\%)} &
\textbf{(\%)} &
\textbf{(ns)} &
\textbf{(\%)}
\\
\hline

P0S0    & 309 & 16.0 & 40.5 & 294.0 & 4.6 & 22.8 & $4.10\pm0.01$ & -- \\

P04S020 & 405 & 32.3 & 13.4 & 419.6 & 69 & 64.4 & $1.62\pm0.01$ & -- \\
P08S016 & 404 & 26.9 & 13.4 & 419.5 & 64 & 59.8 & $1.59\pm0.01$ & -- \\
P08S020 & 405 & 26.6 & 13.3 & 419.6 & 57 & 52.6 & $1.61\pm0.01$ & -- \\

P12S016 & 404 & 27.3 & 13.3 & 419.6 & 63 & 53.6 & $1.61\pm0.01$ & 40 \\
P12S020 & 405 & 34.2 & 13.4 & 419.3 & 69 & 60.1 & $1.62\pm0.01$ & -- \\

P16S016 & 405 & 36.1 & 13.3 & 419.6 & 76 & 67.4 & $1.63\pm0.01$ & 45 \\
P16S020 & 405 & 30.7 & 13.4 & 420.1 & 66 & 59.7 & $1.62\pm0.01$ & -- \\

P20S016 & 404 & 32.7 & 13.3 & 420.0 & 78 & 68.9 & $1.62\pm0.01$ & 52 \\
P20S020 & 405 & 37.1 & 13.3 & 419.9 & 85 & 71.2 & $1.65\pm0.01$ & 51 \\

P30S018 & 405 & 25.6 & 13.4 & 419.7 & 58 & 48.6   & $1.65\pm0.01$ & 66 \\
P30S024 & 406 & 22.6 & 13.6 & 420.3 & 50 & 43.7 & $1.60\pm0.01$ & 70 \\
P30S030 & 407 & 34.1 & 13.7 & 420.9 & 74 & 66.0 & $1.69\pm0.01$ & 70 \\
P30S036 & 407 & 32.0 & 13.7 & 421.3 & 76 & 70.6 & $1.68\pm0.01$ & 71 \\
P30S042 & 408 & 35.8 & 13.8 & 421.3 & 75 & 69.4 & $1.69\pm0.01$ & 66 \\

P40S024 & 405 & 25.7 & 13.5 & 420.1 & 61 & 56.8 & $1.62\pm0.01$ & 78 \\
P40S032 & 407 & 27.7 & 13.6 & 420.8 & 60 & 57.2 & $1.67\pm0.01$ & 85 \\
P40S040 & 408 & 28.1 & 13.9 & 421.3 & 62 & 56.9 & $1.73\pm0.01$ & 81 \\
P40S048 & 408 & 19.8 & 13.9 & 421.3 & 47 & 45.2 & $1.70\pm0.01$ & 81 \\
P40S056 & 409 & 16.9 & 14.1 & 421.6 & 36 & 34.4 & $1.74\pm0.01$ & 73 \\

\textbf{P50S030} &
\textbf{406} &
\textbf{25.9} &
\textbf{13.5} &
\textbf{420.6} &
\textbf{60} &
\textbf{53.8} &
$\mathbf{1.62\pm0.01}$ &
\textbf{100} \\

P50S040 & 408 & 28.6 & 13.8 & 421.2 & 58 & 54.9 & $1.67\pm0.01$ & 92 \\
P50S050 & 408 & 25.3 & 13.9 & 422.1 & 58 & 55.7 & $1.72\pm0.01$ & 91 \\
P50S060 & 410 & 21.7 & 14.2 & 422.1 & 43 & 40.4 & $1.75\pm0.01$ & 97 \\
P50S070 & 410 & 26.2 & 14.1 & 422.5 & 54 & 53.6 & $1.69\pm0.01$ & 91 \\

\hline
\end{tabular}%
}
\end{table}

\subsection{MIP-based Light Output}
\label{subsec:Cosmics_Res}

\begin{figure}[h!]
\centering
\begin{subfigure}[t]{0.48\linewidth}
  \centering
  \includegraphics[width=\linewidth]{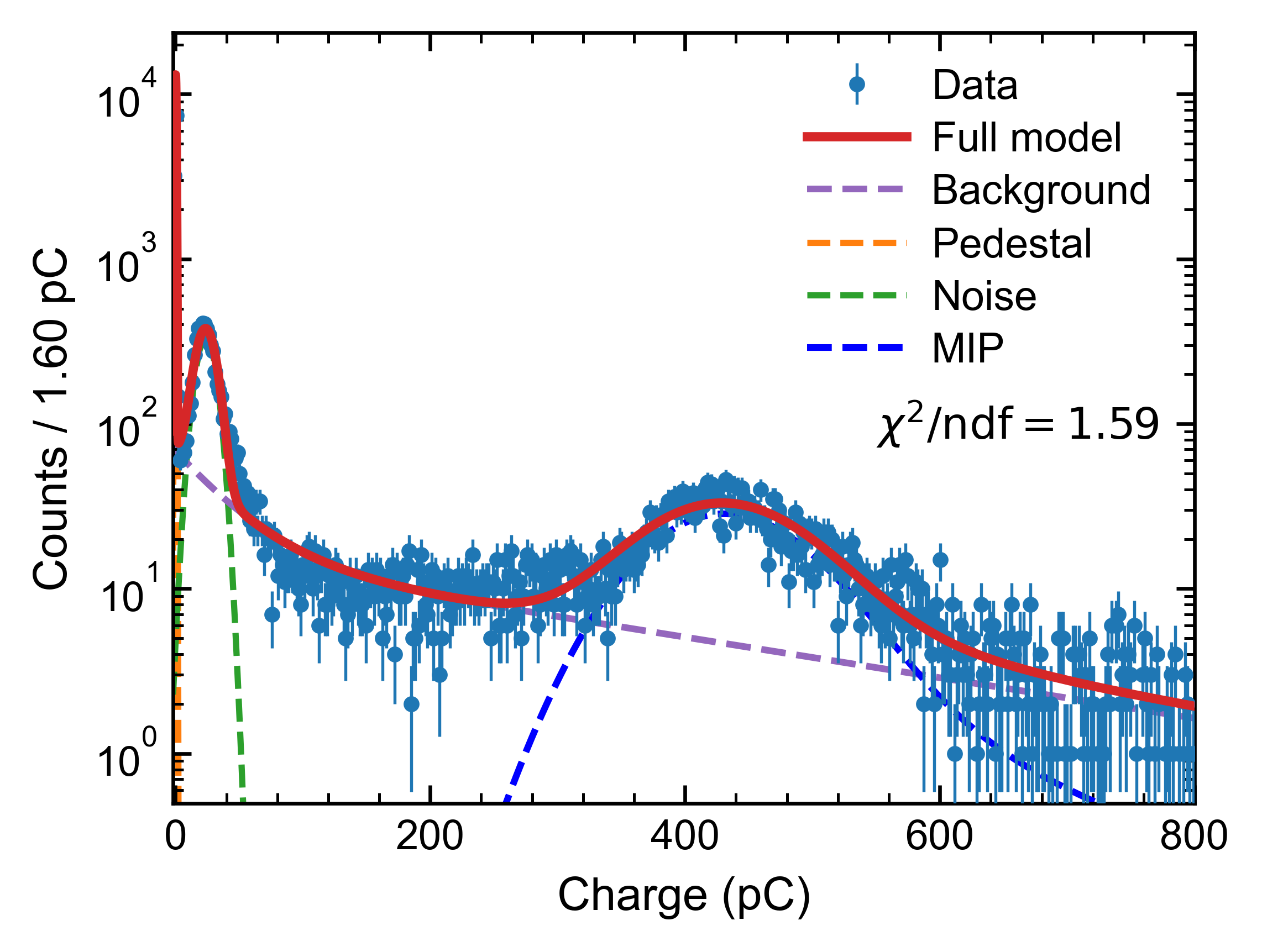}
  \caption{MIP position of the sample P50S030}
  \label{fig:MIP_q}
\end{subfigure}\hfill
\begin{subfigure}[t]{0.48\linewidth}
  \centering
  \includegraphics[width=\linewidth]{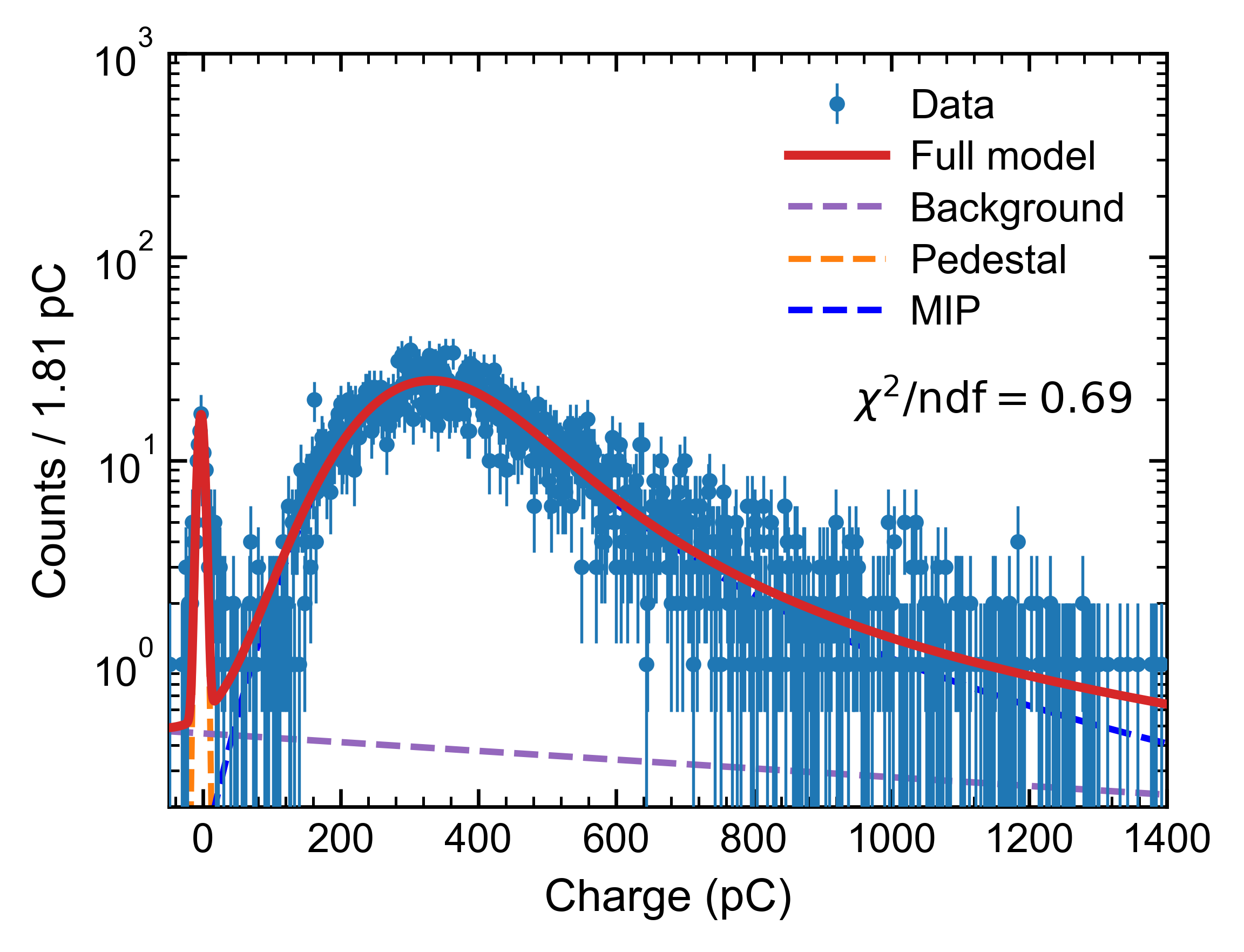}
  \caption{MIP position of the 1~m bar with the trigger at 80~$cm$}
  \label{fig:MIP_bar_1m_80cm}
\end{subfigure}
\caption{MIP charge spectra for the reference sample (left) and for a 1~m scintillator bar (right).}
\label{fig:MIP_sample_and_bar}
\end{figure}

The MIP response of each sample was determined from the measured charge distribution by fitting the model defined in Eq.~\ref{eq:MIP_q}. Representative charge distributions and fits are shown in Fig.~\ref{fig:MIP_sample_and_bar}. The extracted MIP charge, $Q_{\rm MIP}$, was normalized to the sample thickness to account for
differences in the traversed scintillator thickness. Among the investigated samples, P50S030 exhibited the largest thickness-normalized MIP response and was therefore used as the reference for the relative light output. The resulting relative light-output values are summarized in Table~\ref{tab:SampleProperties}.

The stability of the noise component was also examined across the measurements. The fitted noise centroid, $\mu_{\rm noise}$, remained within 22.5--24.5~pC for all samples, while a control measurement performed without a scintillator sample yielded
$\mu_{\rm noise}=23.3\pm0.8$~pC. The width of the noise component was
also relatively stable, with an average $\sigma_{\rm noise}\approx11.7$~pC. The agreement between the sample and no-sample measurements indicates that this contribution is independent of the scintillator and supports its assignment to the
PMT/electronic response.

The dependence of the MIP response on fluor composition is shown in Fig.~\ref{fig:MIP_q_POPOP}. Comparison with the compositions listed in Table~\ref{tab:Samples} reveals a pronounced increase in the MIP response with PPO concentration over the investigated range. This trend is particularly evident when samples with comparable POPOP concentrations are compared while varying the primary-fluor loading.

In contrast, varying the POPOP concentration at fixed PPO loading produces a weaker and non-monotonic dependence. Within the P30 to P50 series, increasing the POPOP/PPO ratio does not systematically increase the MIP response, and the highest POPOP loading does not correspond to the highest response in any of these series. The dominant composition dependant trend is therefore associated with PPO loading, while the secondary-fluor concentration produces smaller variations around this overall dependence.

The MIP response increases substantially with PPO concentration over the investigated range, with the highest response obtained at 5~wt\% PPO. Since higher PPO concentrations were not investigated, 5~wt\% should not be regarded as an absolute optimum, but rather as the upper boundary of the present composition range. Within this range, a high MIP response was achieved while maintaining a fast temporal response. Further increases in PPO may provide additional gains, but may also introduce concentration-dependent losses such as self-absorption and concentration quenching.

From the cosmic-ray measurements, the 200~$mm$ scintillator bar yielded
$37.5\pm0.5$~phe per MIP with the trigger positioned at its center.
For the 1000~$mm$ bar, the measured response decreased with increasing
distance from the SiPM, from $33.5\pm0.2$~phe per MIP at 200~$mm$ to
$23.1\pm0.2$~phe per MIP at 500~$mm$ and $21.2\pm0.2$~phe per MIP at
800~$mm$. Thus, over a propagation distance increase from 200 to 800~$mm$,
the detected MIP response decreased by approximately 37\%. This behavior
is consistent with attenuation during light transport along the bar,
including absorption and losses associated with photon propagation and
surface reflections. Fig.~\ref{fig:MIP_bar_1m_80cm} shows the charge
distribution obtained for the most distant measurement position,
800~$mm$ from the SiPM.

Despite the reduction in collected light with propagation distance, the
MIP contribution remained clearly distinguishable from the pedestal
over the full bar length. The separation was approximately
$5.9\,\sigma_{\rm MIP}$ for the 200~$mm$ bar and
$5.2\,\sigma_{\rm MIP}$, $4.5\,\sigma_{\rm MIP}$, and
$4.3\,\sigma_{\rm MIP}$ for the 1000~$mm$ bar at distances of 200, 500,
and 800~$mm$ from the SiPM, respectively. The simultaneous decrease in
photoelectron yield and MIP--pedestal separation with distance is
consistent with the reduced number of scintillation photons reaching
the SiPM. Nevertheless, even at 800~$mm$, the measured response of
$21.2\pm0.2$~phe per MIP and a separation of approximately
$4.3\,\sigma_{\rm MIP}$ demonstrate that a detectable MIP signal is
maintained over a substantial propagation distance. The results are
summarized in Table~\ref{tab:bar_MIP_performance}.
\begin{table*}[htbp]
\centering
\small
\setlength{\tabcolsep}{5pt}
\renewcommand{\arraystretch}{1.50}
\setlength{\extrarowheight}{5pt}

\caption{Comparison of the performance of the scintillator bars investigated
in this work with representative plastic-scintillator detectors reported
in the literature.}
\label{tab:bar_MIP_performance}

\begin{tabularx}{\textwidth}{
>{\centering\arraybackslash}p{2.55cm}
>{\centering\arraybackslash}p{2.20cm}
>{\centering\arraybackslash}p{1.70cm}
>{\centering\arraybackslash}X
>{\centering\arraybackslash}X
}
\hline

\textbf{Material / Ref.} &
\textbf{Geometry (mm$^3$)} &
\textbf{Readout} &
\textbf{MIP response} &
\textbf{Other performance} \\

\hline


\multirow{4}{*}{\textbf{This work}} &
$200\times50\times10$ &
\shortstack[c]{WLS\\SiPM} &
$37.5\pm0.5$ phe/MIP (100 mm) &
$S_{\rm MIP-ped}=5.9\,\sigma_{\rm MIP}$ \\
\cline{2-5}

&
\multirow{3}{*}{$1000\times50\times10$} &
\multirow{3}{*}{\shortstack[c]{WLS\\SiPM}} &
$33.5\pm0.2$ phe/MIP (200 mm) &
$S_{\rm MIP-ped}=5.2\,\sigma_{\rm MIP}$ \\

&
&
&
$23.1\pm0.2$ phe/MIP (500 mm) &
$S_{\rm MIP-ped}=4.5\,\sigma_{\rm MIP}$ \\

&
&
&
$21.2\pm0.2$ phe/MIP (800 mm) &
$S_{\rm MIP-ped}=4.3\,\sigma_{\rm MIP}$ \\

\hline


\shortstack[c]{FNAL-NICADD\\\cite{Alfaro_2024}}  &
$1000\times40\times10$   &
\shortstack[c]{WLS\\SiPM} &
40 phe/muon &
$\sigma_t=1.7$ ns\\

\hline


\shortstack[c]{Plastic scintillator\\\cite{Sharan2021}} &
$400\times200\times10$  &
SiPM &
22 phe/muon &
$\sigma_t=1.3$ ns \\

\hline


\shortstack[c]{Plastic scintillator\\\cite{Denisov2016}} &
--  &
\shortstack[c]{WLS\\SiPM} &
up to 36 phe/muon/SiPM &
\shortstack[c]{$\sigma_t=0.45$ ns\\$\sigma_x=7.7$ $mm$} \\

\hline


\multirow{2}{*}{
    \shortstack[c]{Clear PS scintillator\\\cite{Denisov2017}}
} &
$1000\times40\times10$ &
\shortstack[c]{4 WLS\\2 SiPM} &
\shortstack[c]{$46\pm2$ / $64\pm7$\\phe/muon/strip (LG/HG)} &
$\lambda_{\rm att}=2.8\pm0.3$ m \\
\cline{2-5}

&
$2000\times40\times10$ &
\shortstack[c]{4 WLS\\2 SiPM} &
\shortstack[c]{$35\pm1$ / $49\pm5$\\phe/muon/strip (LG/HG)} &
$\lambda_{\rm att}=3.3\pm0.2$ m \\

\hline


\shortstack[c]{Bicron R404A\\\cite{Denisov2017}} &
$1000\times27\times12$ &
\shortstack[c]{7 WLS\\2 SiPM} &
\shortstack[c]{$98\pm5$ / $137\pm14$\\phe/muon/strip (LG/HG)} &
\shortstack[c]{
$\lambda_{\rm att}=3.1\pm0.5$ m\\
$\sigma_t\simeq0.33$ ns\\
$\sigma_x\simeq5.4$ cm} \\

\hline


\shortstack[c]{PS + 2\% pTP\\+ 0.03\% POPOP\\\cite{Artikov2016}} &
$2000\times33\times17^{a}$ &
\shortstack[c]{WLS\\PMT} &
Relative light yield &
\shortstack[c]{Optical filler:\\$1.6$--$1.9\times$ gain} \\

\hline


\multirow{2}{*}{
    \shortstack[c]{PS-based scintillator\\\cite{Luo2025}}
} &
$1500\times200\times20$ &
\shortstack[c]{WLS\\4 SiPM/end} &
\shortstack[c]{$\geq51.5$ phe/end\\(750 mm)} &
$\varepsilon_{\mu}>99.3\%$ \\
\cline{2-5}

&
$2000\times200\times20$ &
\shortstack[c]{WLS\\4 SiPM/end} &
\shortstack[c]{$\geq40.8$ phe/end\\(1000 mm)} &
$\varepsilon_{\mu}>99.3\%$ \\

\hline


\shortstack[c]{EJ-200 (PVT)\\\cite{Bloise2023}} &
$1600\times25\times15$ &
\shortstack[c]{2 SiPM} &
-- &
BAL $\simeq3.8$ m \\

\hline


\shortstack[c]{Plastic scintillator\\\cite{Artikov2024}} &
$3000\times40\times7$ &
\shortstack[c]{1 WLS\\SiPM} &
-- &
\shortstack[c]{Design requirement:\\
$\varepsilon_{\mu}\geq99.99\%$} \\

\hline

\end{tabularx}

\vspace{3pt}

\begin{flushleft}
\footnotesize
Geometry is given as length $\times$ width $\times$ thickness.
LG and HG denote the low- and high-gain measurements, respectively.
$S_{\rm MIP-ped}$ denotes the separation between the MIP and pedestal
distributions.
BAL denotes the bulk attenuation length.
$^{a}$ Triangular cross section.
\end{flushleft}

\end{table*}

The detector-scale results are compared in Table~\ref{tab:bar_MIP_performance} with representative plastic-scintillator systems. The comparison is intended to establish the scale of the detected MIP response rather than to rank intrinsic scintillator light output, since the measured photoelectron signal also depends on detector geometry, propagation distance, WLS-fiber configuration, optical coupling, photosensor arrangement, and response definition. 
The main benchmark to compare is the bar studied reported at Alfaro et al.~\cite{Alfaro_2024}, where they reported to collect 40 phe per MIP in a very similar configuration.
Among other systems, the clear-polystyrene scintillator strips of Denisov et al.~\cite{Denisov2017} provide the close comparison, with a $40\times10$~mm$^2$ cross section, lengths of 1--2~m, four WLS fibers, dual-ended SiPM readout, and responses of several tens of photoelectrons per muon. The present $50\times10$~mm$^2$ bar similarly operates in the tens-of-photoelectrons regime.

The broader comparison demonstrates why the complete optical chain must be considered when interpreting absolute photoelectron values. Larger responses obtained with the seven-WLS-fiber Bicron R404A configuration \cite{Denisov2017} and the TAO modules \cite{Luo2025} involve different light-collection architectures and response definitions. Artikov et al.~\cite{Artikov2016} further demonstrated that modification of the optical coupling between the scintillator and WLS fiber can increase the collected light by approximately 1.6--1.9 times. At larger scales, light transport and particle-detection requirements become increasingly relevant: Bloise et al.~\cite{Bloise2023} reported a bulk attenuation length of approximately 3.8~m for EJ-200 bars, while Artikov et al.~\cite{Artikov2024} investigated meter-scale WLS-fiber/SiPM strips for systems requiring muon-detection efficiencies of $99.99\%$ or higher. These results emphasize that the photoelectron response is a detector-level observable rather than a material-specific light-yield benchmark.

\subsection{Radiation Hardness}

\begin{figure}[H]
  \centering
  \begin{subfigure}[t]{0.48\textwidth}
    \centering
    \includegraphics[width=\linewidth]{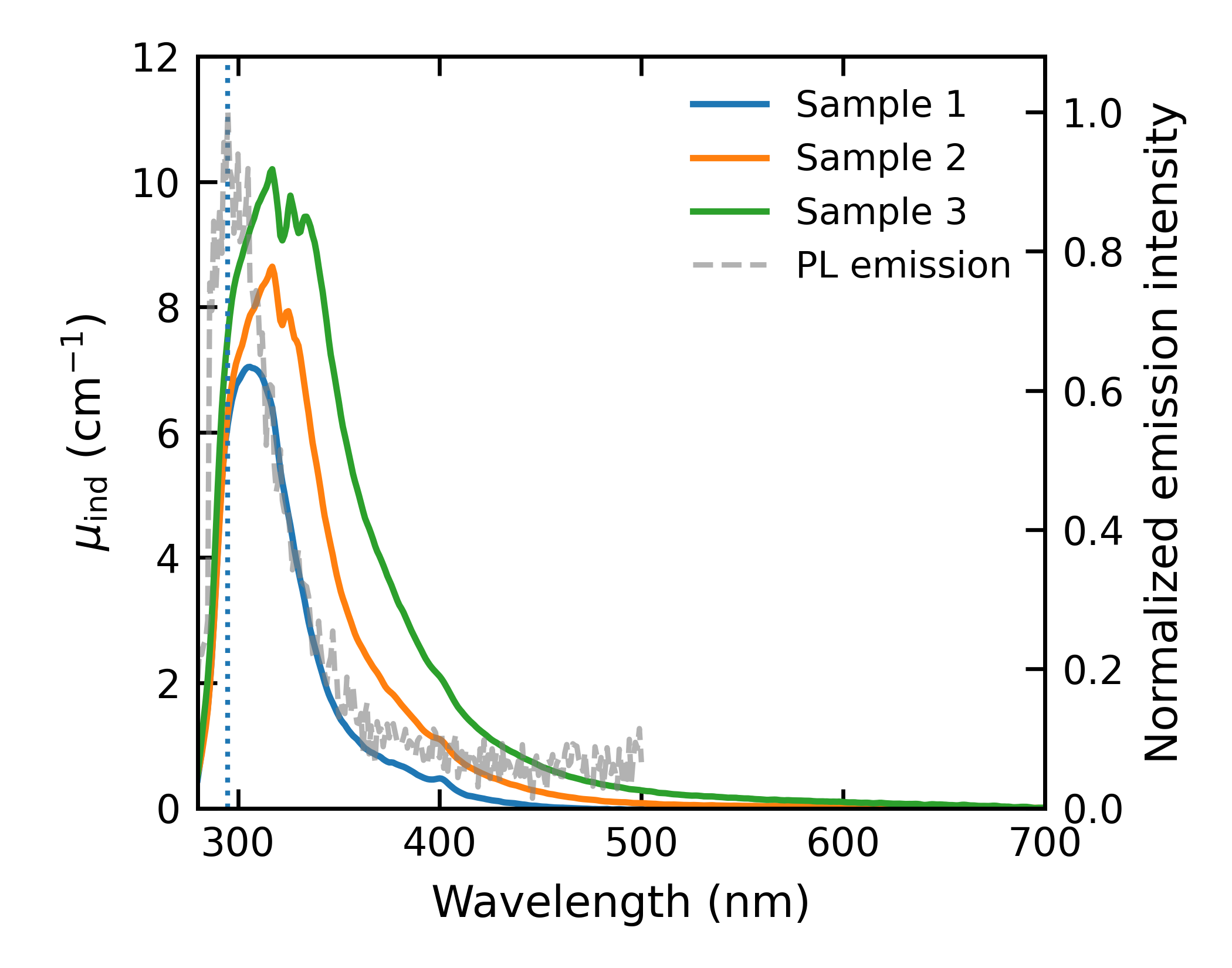}
    \caption{P0S0 samples}
    \label{fig:p0s0_T}
  \end{subfigure}\hfill
  \begin{subfigure}[t]{0.48\textwidth}
    \centering
    \includegraphics[width=\linewidth]{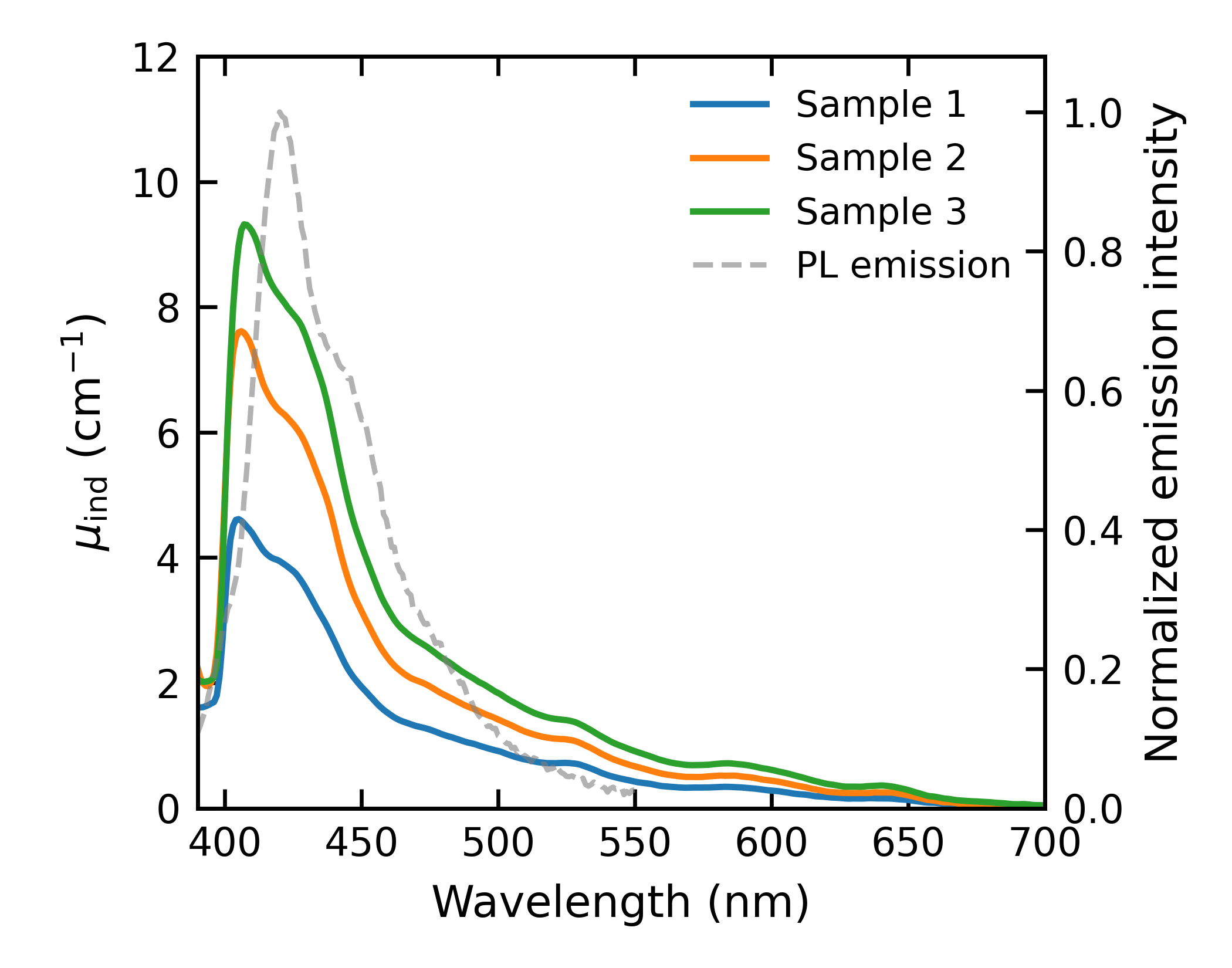}
    \caption{P50S030 samples}
    \label{fig:p50s030_T}
  \end{subfigure}

  \caption{Photo luminescence emission and the induced absorption coefficient spectra before and after irradiation for the (a) P0S0 and (b) P50S030 samples.}
  \label{fig:irrad_T}
\end{figure}

The induced absorption spectra before and after irradiation are shown in Fig.~\ref{fig:irrad_T}, calculated using equation~\ref{eq:induced_absorption}, for the P0S0 and P50S030 batches. The transmission measurements provide direct evidence that proton irradiation introduces additional optical attenuation in P50S030. Its induced absorption coefficient increased systematically with irradiation level. The strongest induced absorption was observed at 405--407~nm for all three irradiated samples, indicating that the spectral position of the dominant damage band remained nearly unchanged while its magnitude increased with exposure.

Importantly, this induced-absorption band strongly overlaps with the POPOP emission region. At the measured POPOP emission maximum of 420~nm, $\mu_{\rm ind}$ increased from 3.94~$cm$$^{-1}$ for Sample~1 to 6.35~$cm$$^{-1}$ for Sample~2 and 8.17~$cm$$^{-1}$ for Sample~3. When the induced absorption was weighted by the pristine P50S030 emission spectrum, the corresponding effective values, $\langle\mu_{\rm ind}\rangle_{\rm em}$, increased from 2.77 to 4.51 and 5.77~$cm$$^{-1}$, respectively. Thus, the irradiation-induced optical degradation occurs directly within the wavelength region in which a substantial fraction of the POPOP fluorescence is emitted.

This spectral overlap is particularly relevant to the scintillation response because POPOP photons generated within the bulk material must propagate through the scintillator before reaching the optical readout. The increasing induced absorption in the POPOP emission region therefore provides an additional attenuation mechanism beyond the reduction of the fluor photoluminescence itself. This is consistent with the substantially larger loss observed in the scintillation light output than in the PL measurements. However, the present measurements do not identify the chemical origin of the induced-absorption band; it may involve radiation-induced changes in the polymer matrix, the fluor system, or their combined optical response.

\begin{figure}[H]
  \centering
  \begin{subfigure}[t]{0.32\textwidth}
    \centering
    \includegraphics[width=\linewidth]{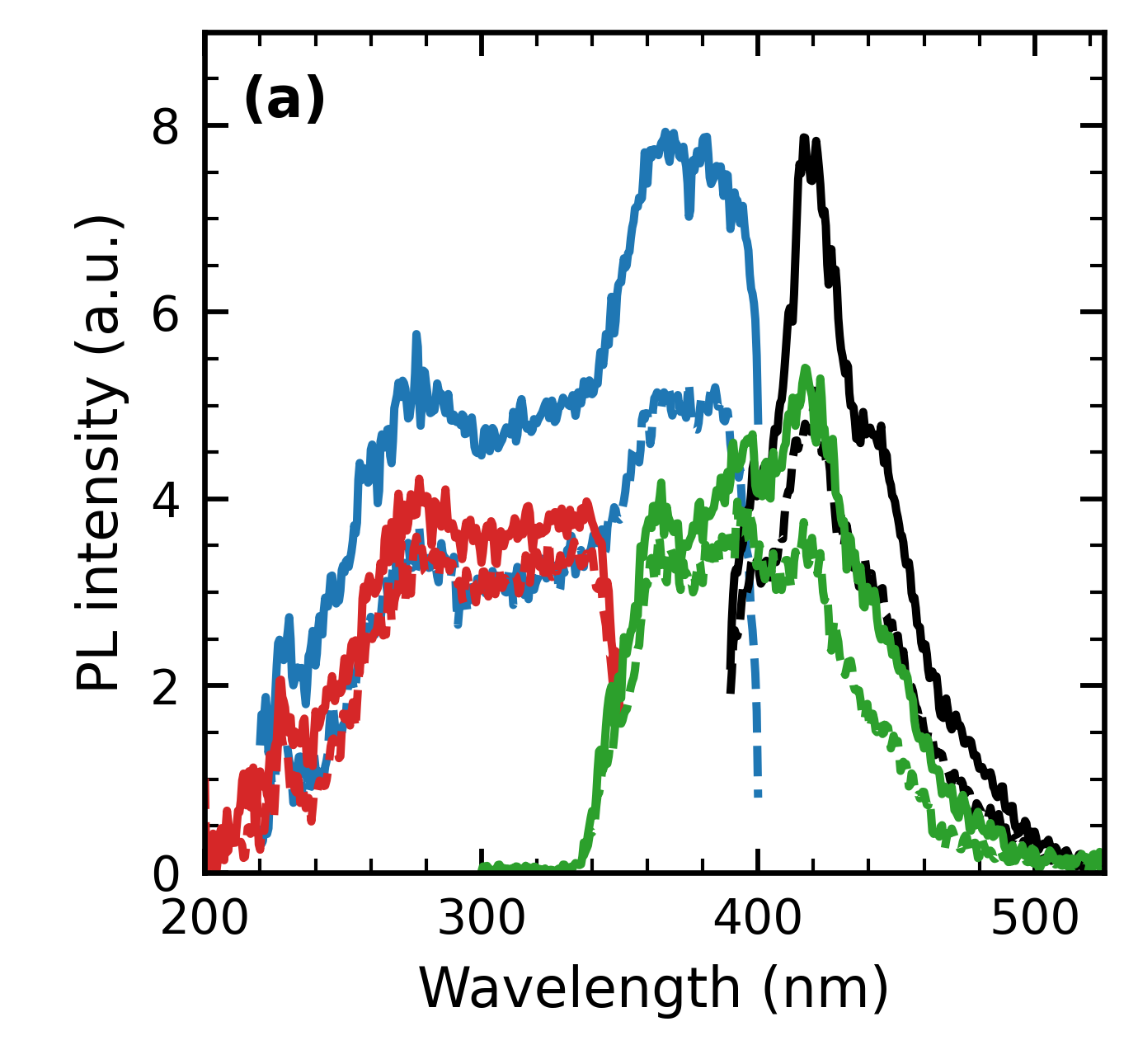}
    \caption{Sample 1}
    \label{fig:irrad_PL_S1}
  \end{subfigure}\hfill
  \begin{subfigure}[t]{0.32\textwidth}
    \centering
    \includegraphics[width=\linewidth]{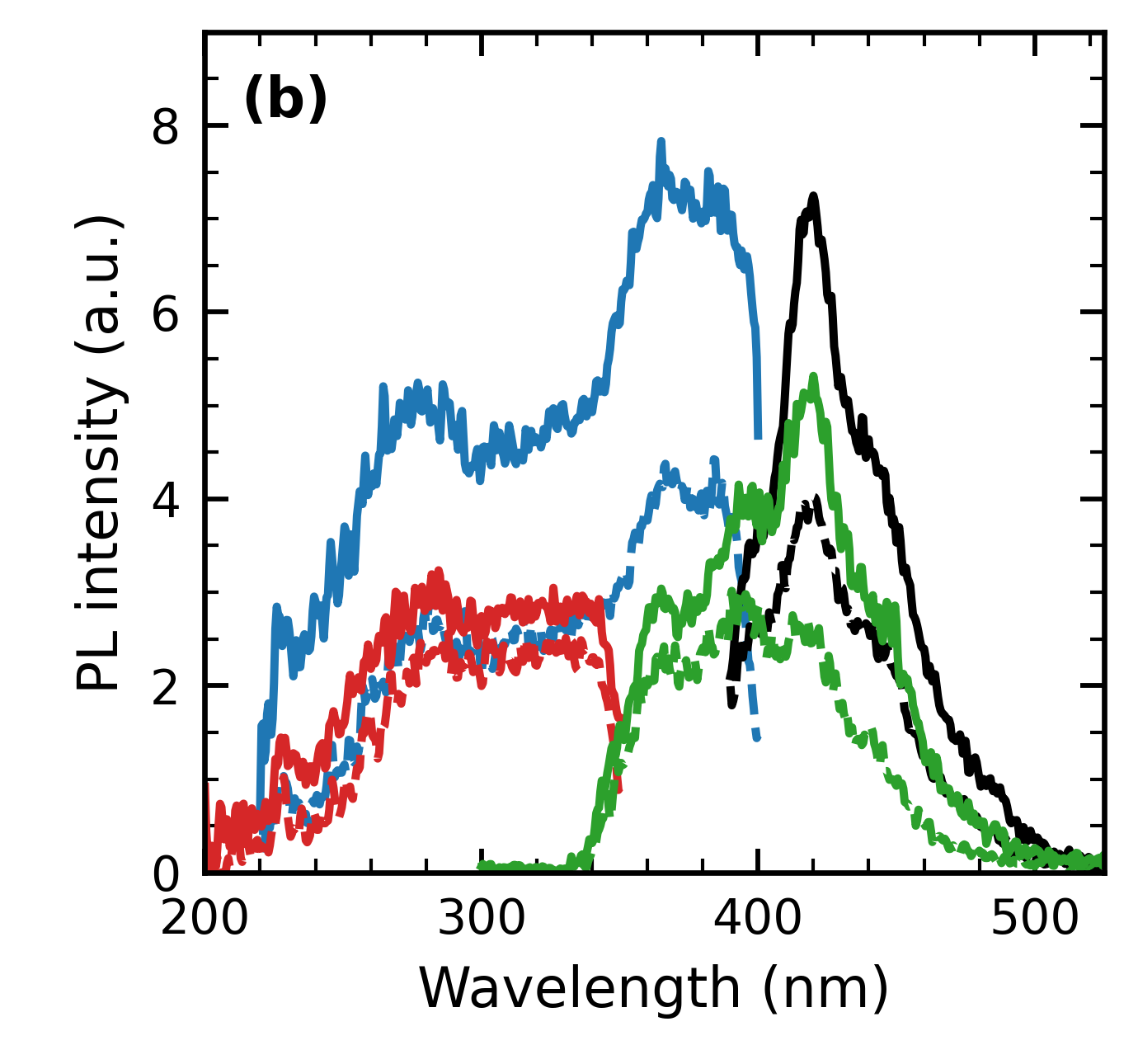}
    \caption{Sample 2}
    \label{fig:irrad_PL_S2}
  \end{subfigure}\hfill
  \begin{subfigure}[t]{0.32\textwidth}
    \centering
    \includegraphics[width=\linewidth]{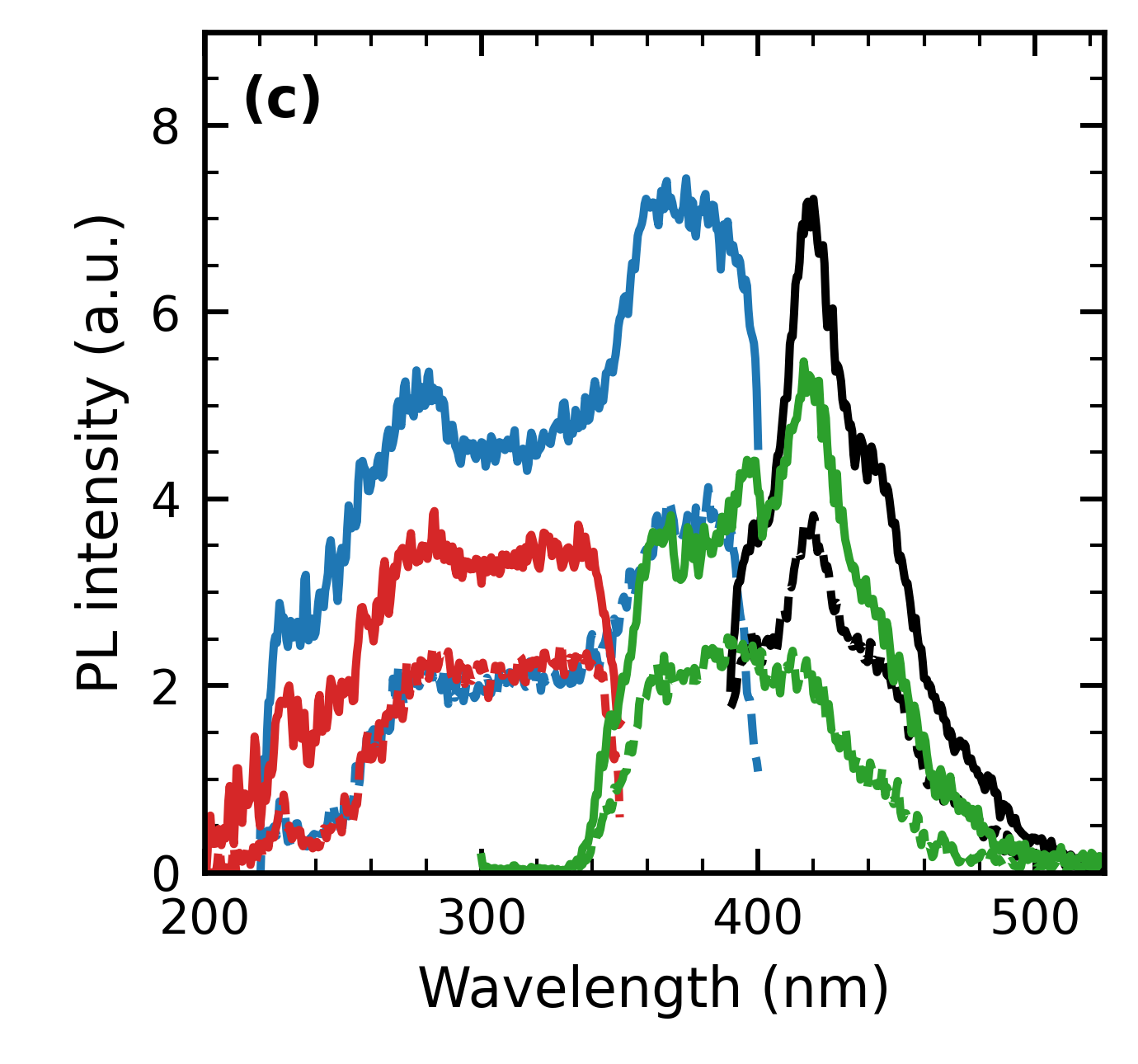}
    \caption{Sample 3}
    \label{fig:irrad_PL_S3}
  \end{subfigure}
  \caption{Photoluminescence changes under proton irradiation. The blue lines are the excitation spectra for an emission wavelength of 418~nm. The red lines are the excitation spectra for an emission wavelength of 378~nm. The black lines are the emission spectra for an excitation wavelength of 378~nm. The green lines are the emission spectra for an excitation wavelength of 280~nm. Solid lines correspond to measurements before irradiation, while dashed lines correspond to measurements after irradiation.}
  \label{fig:irrad_PL}
\end{figure}

The irradiation-induced changes in the photoluminescence response increase systematically from the low- to the high-exposure samples, as can be seen in Fig.~\ref{fig:irrad_PL} and Table~\ref{tab:irrad_summary}. The excitation spectra monitored at the PPO emission peak ($\lambda_{\rm em}=378$~nm) exhibit integral reductions of 18\%, 27\%, and 44\%, respectively, whereas larger reductions of 38\%, 50\%, and 58\% are observed when monitoring at the POPOP emission peak ($\lambda_{\rm em}=418$~nm). Thus, throughout the investigated irradiation series, the excitation pathways feeding the POPOP emission are affected more strongly than those contributing to PPO emission.

The emission measurements show consistent degradation of the fluor response. Under excitation at 280~nm, where the measured spectrum contains contributions from the PPO--POPOP excitation/emission cascade, the integrated intensity decreases by 27\%, 40\%, and 52\% from the low- to the high-exposure samples. Under excitation at 378~nm, predominantly probing the POPOP emission, the corresponding reductions are 33\%, 41\%, and 45\%. The different changes observed for the PPO- and POPOP-associated excitation and emission responses demonstrate that irradiation does not simply scale the fluorescence intensity uniformly. Instead, the fluor system is modified in a pathway-dependent manner. In particular, the preferential suppression of the excitation pathways feeding POPOP emission is consistent with a weakening of the effective PPO-to-POPOP energy-transfer pathway.

\begin{table}[htbp]
\centering
\footnotesize
\setlength{\tabcolsep}{3.2pt}
\renewcommand{\arraystretch}{1.20}

\caption{Radiation-induced changes in the optical and scintillation
response of P50S030.}
\label{tab:irrad_summary}

\begin{tabularx}{\columnwidth}{
>{\raggedright\arraybackslash}p{0.29\columnwidth}
>{\raggedright\arraybackslash}X
>{\centering\arraybackslash}p{0.10\columnwidth}
>{\centering\arraybackslash}p{0.13\columnwidth}
>{\centering\arraybackslash}p{0.10\columnwidth}}
\hline

\textbf{Observable} &
\textbf{Probed response} &
\textbf{Sample 1} &
\textbf{Sample 2} &
\textbf{Sample 3} \\
\hline

\multicolumn{5}{l}{\textit{Photoluminescence integral reduction (\%)}} \\[1pt]

Excitation,
$\lambda_{\rm em}=378$ nm &
PPO emission channel &
18 & 27 & 44 \\

Excitation,
$\lambda_{\rm em}=418$ nm &
POPOP emission channel &
38 & 50 & 58 \\

Emission,
$\lambda_{\rm ex}=280$ nm &
PPO--POPOP excitation/emission cascade &
27 & 40 & 52 \\

Emission,
$\lambda_{\rm ex}=378$ nm &
Predominantly POPOP emission &
33 & 41 & 45 \\

\hline

\multicolumn{5}{l}{\textit{Radiation-induced absorption}} \\[1pt]

$\lambda_{\max}^{\mu_{\rm ind}}$ (nm) &
Maximum induced-absorption wavelength &
405 & 406 & 407 \\

$\mu_{\rm ind}(420~{\rm nm})$ ($cm$$^{-1}$) &
At POPOP emission maximum &
3.94 & 6.35 & 8.17 \\

$\langle\mu_{\rm ind}\rangle_{\rm em}$ ($cm$$^{-1}$) &
POPOP-emission-weighted induced absorption &
2.77 & 4.51 & 5.77 \\

\hline

\multicolumn{5}{l}{\textit{Scintillation light-output retention (\%)}} \\[1pt]

Immediately after irradiation &
Compton-edge response &
22 & 16 & 13 \\

After 2 months recovery &
Compton-edge response &
69 & 50 & 34 \\

\hline
\end{tabularx}
\end{table}

\begin{figure}[H]
  \centering
  \includegraphics[width=0.8\textwidth]{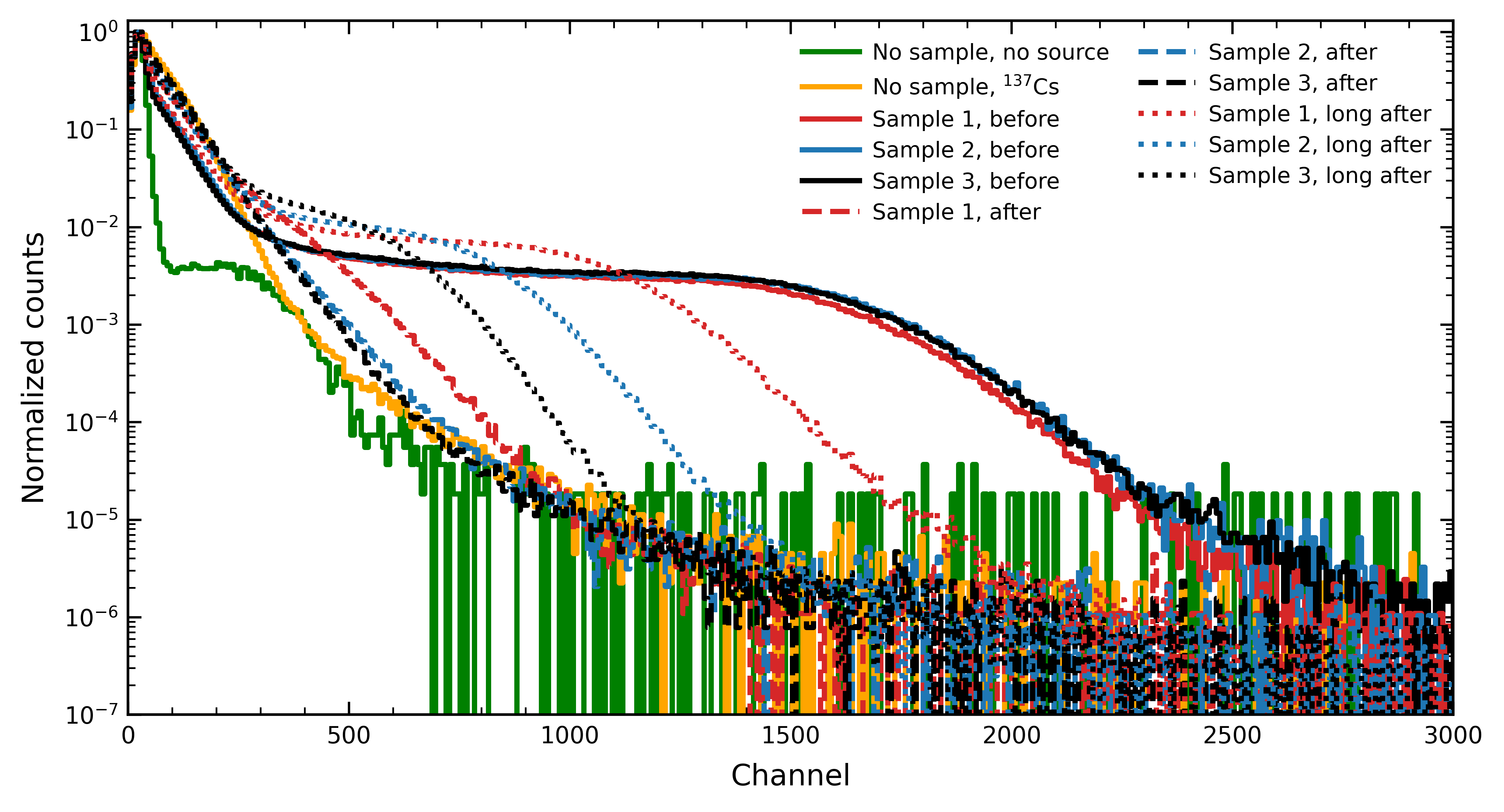}
  
  \caption{Charge spectra of the P50S030 samples measured with a $^{137}$Cs source before and after proton irradiation. Samples 1, 2, and 3 (low, intermediate, and high exposure) are shown in red, blue, and black, respectively. Solid, dashed, and dotted lines correspond to measurements before irradiation, immediately after irradiation, and after approximately two months of recovery, respectively. The green line shows the PMT noise without sample and source; and the yellow line the response to the $^{137}$Cs source without sample.}
\label{fig:irrad_LY}

\end{figure}

The degradation is considerably stronger in the scintillation light-output measurements, as shown in Fig.~\ref{fig:irrad_LY} and Table~\ref{tab:irrad_summary}. Immediately after irradiation, the Compton-edge response retains only 22\%, 16\%, and 13\% of the pre-irradiation response for Samples~1, 2, and 3, respectively. These losses are substantially larger than the corresponding changes observed in the PL measurements, indicating that fluor-emission degradation alone cannot account for the reduction in detected scintillation light. This interpretation is further supported by the radiation-induced absorption measurements. The dominant induced-absorption feature occurs at 405--407~nm, overlapping the POPOP emission region, while $\mu_{\rm ind}$ at the measured POPOP emission maximum (420~nm) increases from 3.94 to 6.35 and 8.17~$cm$$^{-1}$ for Samples~1,2 and 3, respectively. The corresponding emission-weighted induced absorption, $\langle\mu_{\rm ind}\rangle_{\rm em}$, increases from 2.77 to 4.51 and 5.77~$cm$$^{-1}$. The combined PL, transmission, and scintillation measurements therefore indicate that the severe loss of light output involves both modifications of the excitation/emission processes and increased attenuation of POPOP-region photons during optical transport.

A pronounced light-output recovery is observed after approximately two months. The retained light output increases from 22\% to 69\%, from 16\% to 50\%, and from 13\% to 34\% for the low-, intermediate-, and high-exposure samples, respectively. Irradiation, therefore, produces a substantial recoverable component. Nevertheless, the incomplete recovery and the systematic decrease of the recovered response with increasing exposure demonstrate the accumulation of persistent radiation-induced degradation.
 
The distinction between the immediate and recovered light output also provides an approximate indication of the exposure scale associated with transient and persistent degradation. For the lowest-exposure sample, the light-output retention was only 22\% immediately after irradiation but recovered to 69\% after approximately two months. Using the nominal irradiation setting only as an approximate exposure scale, linear interpolation between the pristine response and the immediate post-irradiation response gives the estimated dose for 90\% immediate light-output retention $D_{90,\mathrm{imm}}^{\mathrm{lin}}\simeq0.13~\mathrm{Mrad}$. Alternatively, assuming a local exponential response, we obtain $D_{90,\mathrm{imm}}^{\mathrm{exp}}\simeq0.07~\mathrm{Mrad}$. The corresponding estimates based on the recovered response ($L/L_0=0.69$) are approximately 0.32~Mrad and 0.28~Mrad for the linear and exponential parameterizations, respectively. Thus, the nominal exposure associated with a 10\% response reduction is estimated to be on the order of $0.07$--$0.13$~Mrad immediately after irradiation, but increases to approximately $0.28$--$0.32$~Mrad when recovery is taken into account.

These values should be regarded only as order-of-magnitude estimates of the low dose response. In particular, the immediate response over the measured 1--10~Mrad range is strongly non-linear and cannot be described by a single exponential attenuation law. Moreover, no measurements were performed between the pristine condition and the lowest irradiation level, and the absorbed dose in the scintillator was not independently determined. Dedicated measurements in the tens to few hundred kRad range are therefore required to establish the actual onset of both transient and persistent degradation.

\subsection{Discussion}\label{sec:discussion}

The combined measurements indicate that the performance of the PU--PPO--POPOP system is determined by a balance between scintillation conversion, wavelength shifting, and photon transport. Although POPOP dominates the emission of the doped samples, the residual PPO component shows that the spectral conversion is incomplete. Moreover, suppression of PPO emission does not systematically correspond to a higher MIP response. The PPO-to-POPOP spectral ratio should therefore not be interpreted directly as an energy transfer efficiency, since the measured spectra can contain contributions from both non-radiative energy transfer and radiative reabsorption.

The optical and MIP measurements further show that improved spectral conversion does not necessarily improve the detected signal. Increasing the fluor loading modifies both the emission spectrum and the transmission near the emission band, while $T_{\rm eff}$ shows no direct correspondence with the relative MIP response. Within the investigated composition range, increasing PPO produces the dominant improvement in MIP response, whereas varying POPOP at fixed PPO results in smaller and non-monotonic changes. This suggests that the secondary-fluor concentration should be sufficient for wavelength shifting but need not be maximized, because additional loading can also increase concentration-dependent optical losses.

The timing results introduce little additional constraint over the investigated doped compositions since the substantial changes in MIP response occur with only modest variations in $\tau_{\rm eff}$. The sample P50S030 therefore provides the most favorable measured combination of light output and fast response within the investigated composition space. This composition should not be regarded as a global optimum; however, because PPO concentrations above 5~wt\% were not investigated.

Scale-up measurements at meter-scale detectors of 200- and 1000-$mm$ plastic scintillator bars demonstrate that the material performance translates to a WLS-fiber/SiPM detector geometry. Although the detected response decreases with propagation distance, the MIP signal remains distinguishable from the pedestal at 800~$mm$. Its photoelectron response is of the same order as that reported for other WLS fiber plastic-scintillator systems \cite{Denisov2017,Luo2025,Alfaro_2024}. Still, absolute values cannot be compared directly because of differences in geometry, the number of WLS fibers, optical coupling, and readout configuration. Artikov et al.~\cite{Artikov2016} also demonstrate this sensitivity to the complete optical chain. The present results, therefore, support a coupled material–detector optimization strategy in which fluor composition and light-collection architecture are optimized together rather than independently.

In Britvich et al.~\cite{britvich_radiation_1993}, several polystyrene-based formulations for plastic scintillators were irradiated with a dose of 10~Mrad using a $^{137}$Cs source (20~krad/h), followed by 23 days recovery period. Some of these formulations are relevant for the present work, which can be categorized into two main groups based on their dopant concentrations: 1) low-dopant formulations, which are samples with 0.5-2.0\% PPO and 0.01-0.5\% POPOP; and 2) high-dopant formulations with 10\% PPO and 0.5-1.0\% POPOP, with 2 out of the 3 samples in this category containing 5\% naphthalene (N) + 2\% p-terphenyl (pTP) as secondary additives. Formulations in the first category presented a light-output retention of $\sim8-18\%$ immediately after irradiation and recovered to $50-55\%$ after 23 days, which is similar performance obtained in this work (see Table~\ref{tab:irrad_summary}), with retention ranging from 13-22\% and recovery after two months of 34-69\%. In contrast, high-dopant formulations demonstrated significantly greater radiation hardness. The sample containing only PPO and POPOP ratained $\sim37\%$ of its original light output, recovering to $\sim68\%$). While the formulations co-doped with N and pTP had the highest initial retention ($79-83\%$), recovering to $\sim92-97\%$,  we therefore consider this strategy worth exploring to enhance radiation hardness.  

The demonstrated meter-scale performance and radiation response make P50S030 relevant to the ongoing development of scintillator-based detectors for the ALICE~3 Muon Identifier (MID). Its combination of fast timing, efficient MIP detection, scalable bar geometry, and substantial post-irradiation recovery is compatible with the main performance requirements for large-area HEP detectors. Further measurements in the low-dose region, under irradiation conditions representative of the expected MID environment, are needed to quantify its radiation tolerance for this application. Although the MID is our main motivation, other detectors from ALICE and HEP experiments in general may benefit from this type of material.


\section{Conclusion}

The composition dependence of PU--PPO--POPOP plastic scintillators was systematically investigated through optical, timing, and MIP measurements. Within the investigated composition range, increasing the PPO concentration produced the strongest improvement in MIP response, whereas variations in POPOP mainly modified the spectral response and resulted in a weaker, non-monotonic dependence of the light output. The doped samples maintained fast effective decay times of 1.59--1.75~ns.

Among the investigated formulations, P50S030, containing 5.0~wt\% PPO and 0.030~wt\% POPOP, provided the most favorable overall performance. It exhibited the highest measured relative light output, defined as 100\% within the investigated sample set, together with an effective decay time of $1.62\pm0.01$~ns and an emission-weighted transmittance of 53.8\%. These results demonstrate that maximizing optical transmittance alone does not necessarily maximize the detected scintillation response.

The detector-scale measurements further demonstrated the performance of the optimized formulation in a WLS-fiber/SiPM configuration. For the 1-m long bar, the MIP response decreased from $33.5\pm0.2$~phe/MIP at 200~$mm$ to $21.2\pm0.2$~phe/MIP at 800~$mm$ from the SiPM. Nevertheless, the MIP signal remained separated from the pedestal by $4.3\,\sigma_{\rm MIP}$ at 800~$mm$, demonstrating clear MIP identification over the investigated propagation distance.

Proton irradiation produced systematic degradation of both the optical and scintillation responses of P50S030. The PPO- and POPOP-associated PL responses decreased with increasing irradiation level, with the excitation pathways feeding POPOP emission being more strongly affected. The scintillation response exhibited substantially larger degradation than the PL response, indicating that the loss of detected light cannot be attributed to fluor-emission degradation alone and also involves changes in energy-transfer and optical-transport processes. Immediately after irradiation, the retained light output was 22\%, 16\%, and 13\% for the low-, intermediate-, and high-exposure samples, respectively. After approximately two months of recovery, these values increased to 69\%, 50\%, and 34\%, demonstrating a substantial recoverable component together with persistent degradation that increased with irradiation level.
A rough estimate suggests that the low-damage operating regime of P50S030 may extend from several tens to a few hundred krad. Further irradiation measurements in this range are required to quantify the onset of performance degradation and establish a reliable radiation-tolerance limit. 

Overall, P50S030 provides the best balance of light output and fast timing; therefore, it was used for the construction of a large area plastic scintillator and proved to have sufficiently good performance to be considered a candidate for the construction. On the other hand, radiation hardness could be improved by adding dopants that may enhance its resistance.

\section*{Acknowledgements}
The authors gratefully acknowledge the financial support provided by the Mexican Secretary of Science, Humanities, Technology and Education (SECIHTI) through a national doctoral fellowship grant (CVU No. 861419); to the Ministry of Education, Youth and Sports of the Czech Republic project LM2023040.
The authors gratefully acknowledge the CERN PS IRRAD facility and its staff for providing the irradiation infrastructure and technical support. This work was supported by the EU and the Ministry of
Education, Youth and Sports of the Czech Republic under Project No.~CZ.02.01.01/00/22\_011/0008569, ``Czech Technical University -- International Postdoctoral Programme.''

\bibliographystyle{JHEP.bst} 


\bibliography{ref.bib}

@article{schorr_solid_1950,
	title = {Solid {Non}-{Crystalline} {Scintillation} {Phosphors}},
	volume = {80},
	number = {3},
	pages = {474--474},
	journal = {Physical Review},
	author = {Schorr, Marvin G. and Torney, Franklin L.},
	month = nov,
	year = {1950},
	doi = {10.1103/PhysRev.80.474},
	url = {https://link.aps.org/doi/10.1103/PhysRev.80.474},
}

@article{shevelev_optical_2025,
	title = {Optical and scintillation properties of new polyurethane plastic scintillator with {PPO} and {POPOP}},
	journal = {Optical Materials},
	pages = {117202},
	author = {Shevelev, V. S. and Ishchenko, A. V.},
	month = may,
	year = {2025},
	issn = {0925-3467},
	doi = {10.1016/j.optmat.2025.117202},
	url = {https://www.sciencedirect.com/science/article/pii/S0925346725005622},
}

@article{MIZUEHAMADA1999148,
title = {Radiation damage studies on the optical and mechanical properties of plastic scintillators},
journal = {Nuclear Instruments and Methods in Physics Research Section A: Accelerators, Spectrometers, Detectors and Associated Equipment},
volume = {422},
number = {1},
pages = {148-154},
year = {1999},
issn = {0168-9002},
doi = {https://doi.org/10.1016/S0168-9002(98)01091-2},
url = {https://www.sciencedirect.com/science/article/pii/S0168900298010912},
author = {Margarida {Mizue Hamada} and Paulo {Roberto Rela} and Fábio {Eduardo da Costa} and Carlos {Henrique de Mesquita}}
}

@article{Alfaro_2024,
doi = {10.1088/1748-0221/19/04/T04006},
url = {https://doi.org/10.1088/1748-0221/19/04/T04006},
year = {2024},
month = {apr},
publisher = {IOP Publishing},
volume = {19},
number = {04},
pages = {T04006},
author = {Alfaro, Ruben and Alvarado Hernández, Mauricio and Bencédi, Gyula and Cabanillas Noris, Juan Carlos and Díaz Maldonado, Marco Antonio and Duarte Galvan, Carlos and Fernández Téllez, Arturo and Gábor Barnaföldi, Gergely and Gera, Ádám and Grabsky, Varlen and Hamar, Gergő and Herrera Corral, Gerardo and León Monzón, Ildefonso and Martínez García, Josué and Martínez Hernandez, Mario Iván and Muñoz Méndez, Jesús Eduardo and Nagy, Richárd and Narcio Laveaga, Rafael Ángel and Ortiz, Antonio and Rodríguez-Cahuantzi, Mario and Rojas Torres, Solangel and Szollosova, Timea and Patiño Salazar, Miguel Enrique and Pazarán García, Jared and Regules Medel, Hector David and Tejeda Muñoz, Guillermo and Vargas Torres, Paola and Varga, Dezső and Vértesi, Róbert and Vasquez Beltran, Yael Antonio and Vázquez Villamar, Carlos Rafael and Yoval Pozos, Irandheny},
title = {Characterisation of plastic scintillator paddles and lightweight MWPCs for the MID subsystem of ALICE 3},
journal = {Journal of Instrumentation}
}

@article{hu_preparation_2023,
	title = {Preparation and characterization of plastic scintillators with different proportions},
	volume = {332},
	number = {4},
	pages = {1047--1054},
	journal = {Journal of Radioanalytical and Nuclear Chemistry},
	author = {Hu, Yichen and Ma, Jiaju and Liu, Tonghuan and Zhang, Yongjie and Yu, Yuhong and Sun, Zhiyu and Fang, Fang and Dong, Tianhao and Zhao, Yanhong and Zhou, Gen},
	month = apr,
	year = {2023},
	issn = {1588-2780},
	doi = {10.1007/s10967-022-08748-6},
	url = {https://doi.org/10.1007/s10967-022-08748-6},
}

@article{nam_study_2017,
	title = {Study on the {Characteristics} of a {Scintillator} for {Beta}-ray {Detection} using {Epoxy} {Resin}},
	volume = {153},
	pages = {07005},
	journal = {EPJ Web of Conferences},
	author = {Nam, Jong Soo and Choi, Yong Seok and Hong, Sang Bum and Seo, Bum Kyung and Moon, Jei Kwon and Choi, Jong Won},
	month = sep,
	year = {2017},
	doi = {10.1051/epjconf/201715307005},
	url = {https://ui.adsabs.harvard.edu/abs/2017EPJWC.15307005N},
}

@article{shevelev_new_2023,
	title = {New ultrafast low-cost polyurethane based plastic scintillator},
	volume = {263},
	pages = {120133},
	journal = {Journal of Luminescence},
	author = {Shevelev, V. S. and Ishchenko, A. V. and Omelkov, S. I. and Nagirnyi, V.},
	month = nov,
	year = {2023},
	issn = {0022-2313},
	doi = {10.1016/j.jlumin.2023.120133},
	url = {https://www.sciencedirect.com/science/article/pii/S0022231323004660},
}

@article{syntfeld-kazuch_characterization_2024,
	title = {Characterization of a novel polyurethane-based plastic scintillator for neutron and gamma detection in mixed radiation fields},
	volume = {1068},
	pages = {169764},
	journal = {Nuclear Instruments and Methods in Physics Research Section A: Accelerators, Spectrometers, Detectors and Associated Equipment},
	author = {Syntfeld-Każuch, A. and Swiderski, L. and Szczesniak, T. and Grodzicka-Kobylka, M. and Brylew, K. and Mianowska, Z. and Maiatska, O. and Pausch, G. and Stein, J. and Ritter, H.},
	month = nov,
	year = {2024},
	issn = {0168-9002},
	doi = {10.1016/j.nima.2024.169764},
	url = {https://www.sciencedirect.com/science/article/pii/S0168900224006909},
}

@misc{maiatska_polyurethane-based_2026,
	title = {Polyurethane-{Based} {Scintillators} for {Neutron} and {Gamma} {Radiation} {Detection} in {Medical} and {Industrial} {Applications}},
	author = {Maiatska, Olga and Dünnebacke, Torsten and Kreuels, Martin and Pausch, Guntram and Scherwinski, Falko and Stein, Jürgen R.},
	publisher = {arXiv},
	month = jan,
	year = {2026},
	eprint = {2601.17463},
	archivePrefix = {arXiv},
	primaryClass = {physics.ins-det},
	doi = {10.48550/arXiv.2601.17463},
	url = {http://arxiv.org/abs/2601.17463},
}

@article{kral_tailored_2026,
	title = {Tailored thermally stable functionalization of {CsPbBr3} nanocrystals for polymer nanocomposite scintillator fabrication},
	volume = {14},
	number = {11},
	pages = {4440--4450},
	journal = {Journal of Materials Chemistry C},
	author = {Král, Jan and Děcká, Kateřina and Liška, Petr and Torres, Solangel Rojas and Valenta, Jan and Babin, Vladimir and Monzón, Ildefonso León and Čuba, Václav and Mihóková, Eva and Auffray, Etiennette},
	month = mar,
	year = {2026},
	issn = {2050-7534},
	doi = {10.1039/D5TC03614C},
	url = {https://pubs.rsc.org/en/content/articlelanding/2026/tc/d5tc03614c},
}

@article{younesi_effect_2024,
	title = {The effect of {PPO} concentration on the scintillation properties of a polystyrene based plastic scintillator},
	volume = {19},
	number = {01},
	pages = {T01001},
	journal = {Journal of Instrumentation},
	author = {Younesi, L. and Salehi-Barough, M. and Ahmadi, Sh and Esmaili-Sani, V.},
	month = jan,
	year = {2024},
	issn = {1748-0221},
	doi = {10.1088/1748-0221/19/01/T01001},
	url = {https://dx.doi.org/10.1088/1748-0221/19/01/T01001},
}

@article{alex_effect_2022,
	title = {Effect of {PPO} and {POPOP} activators on the scintillation performance of polystyrene-based scintillator},
	volume = {24},
	pages = {365--371},
	journal = {Journal of Optoelectronics and Advanced Materials},
	author = {Alex, Lizbeth and Paulraj, Rajesh and Tyagi, Mohit},
	month = aug,
	year = {2022},
	issn = {1454-4164},
	url = {https://joam.inoe.ro/articles/effect-of-ppo-and-popop-activators-on-the-scintillation-performance-of-polystyrene-based-scintillator/},
}

@article{moore_high-resolution_2025,
	title = {High-resolution {3D}-printed plastic scintillators with tertiary dye},
	volume = {20},
	number = {10},
	pages = {P10043},
	journal = {Journal of Instrumentation},
	author = {Moore, C.J. and Febbraro, M. and Manfredi, J.J. and Wood, A. and Rutstrom, D. and Ruland, T. and Hackett, B. and Hausladen, P.A.},
	month = oct,
	year = {2025},
	issn = {1748-0221},
	doi = {10.1088/1748-0221/20/10/P10043},
	url = {https://doi.org/10.1088/1748-0221/20/10/P10043},
}

@article{Sirunyan_2020,
	title = {Measurements with silicon photomultipliers of dose-rate effects in the radiation damage of plastic scintillator tiles in the {CMS} hadron endcap calorimeter},
	volume = {15},
	number = {06},
	pages = {P06009},
	journal = {Journal of Instrumentation},
	author = {Sirunyan, A.M. and others},
	collaboration = {CMS},
	publisher = {{IOP} Publishing},
	month = jun,
	year = {2020},
	doi = {10.1088/1748-0221/15/06/p06009},
	url = {https://doi.org/10.1088/1748-0221/15/06/p06009},
}

@article{ADAMS2020163970,
	title = {The {STAR} event plane detector},
	volume = {968},
	pages = {163970},
	journal = {Nuclear Instruments and Methods in Physics Research Section A: Accelerators, Spectrometers, Detectors and Associated Equipment},
	author = {Adams, J. and Ewigleben, A. and Garrett, S. and He, W. and Huang, T. and Jacobs, P.M. and Ju, X. and Lisa, M.A. and Lomnitz, M. and Pak, R. and Reed, R. and Schmah, A. and Shanmuganathan, P. and Shao, M. and Sun, X. and Upsal, I. and Visser, G. and Zhang, J.},
	year = {2020},
	issn = {0168-9002},
	doi = {10.1016/j.nima.2020.163970},
	url = {https://www.sciencedirect.com/science/article/pii/S0168900220304344},
}

@article{Belle-II:2010dht,
	title = {{Belle II Technical Design Report}},
	author = {Abe, T. and others},
	collaboration = {Belle-II},
	eprint = {1011.0352},
	archivePrefix = {arXiv},
	primaryClass = {physics.ins-det},
	reportNumber = {KEK-REPORT-2010-1},
	month = nov,
	year = {2010},
}

@article{T2K:2011qtm,
	title = {{The T2K Experiment}},
	author = {Abe, K. and others},
	collaboration = {T2K},
	eprint = {1106.1238},
	archivePrefix = {arXiv},
	primaryClass = {physics.ins-det},
	journal = {Nucl. Instrum. Meth. A},
	volume = {659},
	pages = {106--135},
	year = {2011},
	doi = {10.1016/j.nima.2011.06.067},
}

@article{ALVARADO2020163150,
	title = {A beam–beam monitoring detector for the {MPD} experiment at {NICA}},
	volume = {953},
	pages = {163150},
	journal = {Nuclear Instruments and Methods in Physics Research Section A: Accelerators, Spectrometers, Detectors and Associated Equipment},
	author = {Alvarado, Mauricio and Ayala, Alejandro and Ayala-Torres, Marco Alberto and Bietenholz, Wolfgang and Dominguez, Isabel and Fontaine, Marcos and González-Zamora, P. and Montaño, Luis Manuel and Moreno-Barbosa, E. and Patiño Salazar, Miguel Enrique and Moreno, L.A.P. and Nieto-Marín, P.A. and {Reyna Ortiz}, V.Z. and Rodríguez-Cahuantzi, M. and Tejeda-Muñoz, G. and Tejeda-Yeomans, Maria Elena and Villatoro-Tello, A. and {Zepeda Fernández}, C.H.},
	year = {2020},
	issn = {0168-9002},
	doi = {10.1016/j.nima.2019.163150},
	url = {https://www.sciencedirect.com/science/article/pii/S0168900219314639},
}

@article{Baldini_2017,
	title = {Measurement of parameters of scintillating bars with wavelength-shifting fibres and silicon photomultiplier readout for the {SHiP} Muon Detector},
	volume = {12},
	number = {03},
	pages = {P03005},
	journal = {Journal of Instrumentation},
	author = {Baldini, W. and Blondel, A. and Calcaterra, A. and Jacobsson, R. and Khotjantsev, A. and Kudenko, Y. and Kurochka, V. and Lanfranchi, G. and Mefodiev, A. and Mineev, O. and Montanari, A. and Noah Messomo, E. and Saputi, A. and Tosi, N.},
	publisher = {{IOP} Publishing},
	month = mar,
	year = {2017},
	doi = {10.1088/1748-0221/12/03/p03005},
	url = {https://doi.org/10.1088/1748-0221/12/03/p03005},
}

@misc{polymeres_technologies_100_nodate,
	title = {100\% {UV} {Resistant} {Clear} {Aliphatic} {Polyurethane} {Casting} {Resin} - {TECHNO} {CAST} 3284},
	author = {{Polymeres Technologies}},
	howpublished = {Polymeres Technologies product page},
	url = {https://polymerestechnologies.com/product/uv-resistant-aliphatic-polyurethane-casting-resin/},
	note = {Accessed: 2026-05-18},
}

@article{brosi_time-correlated_2025,
	title = {Time-{Correlated} {Single}-{Photon} {Counting} for versatile longitudinal diagnostics at the {MAX} {IV} {Laboratory} storage rings},
	volume = {20},
	number = {03},
	pages = {P03011},
	journal = {Journal of Instrumentation},
	author = {Brosi, Miriam and Schmand, Johann and Breunlin, Jonas and Curbis, Francesca},
	month = mar,
	year = {2025},
	issn = {1748-0221},
	doi = {10.1088/1748-0221/20/03/P03011},
	url = {https://doi.org/10.1088/1748-0221/20/03/P03011},
}

@article{Gkotse:2896062,
	title = {{CERN Proton Irradiation Facility (IRRAD) Data Management, Control and Monitoring System Infrastructure for post-LS2 Experiments}},
	author = {Gkotse, Blerina and Jouvelot, Pierre and Pezzullo, Giuseppe and Ravotti, Federico},
	journal = {JACoW ICALEPCS},
	volume = {2023},
	pages = {TUPDP093},
	year = {2023},
	doi = {10.18429/JACoW-ICALEPCS2023-TUPDP093},
	url = {https://cds.cern.ch/record/2896062},
}

@article{Sharan2021,
	author = {Sharan, Manoj K. and Singaraju, Ram Narayan and Sinha, Tinku and Ghosh, Tanmoy and Jha, V. N.},
	title = {Measurement of cosmic-ray muon flux in the underground laboratory at {UCIL}, India, using plastic scintillators and {SiPM}},
	journal = {Nuclear Instruments and Methods in Physics Research Section A},
	volume = {994},
	pages = {165083},
	year = {2021},
	doi = {10.1016/j.nima.2021.165083},
}

@article{Denisov2016,
	author = {Denisov, Dmitri and Evdokimov, Valery and Luki{\'c}, Strahinja},
	title = {Time and position resolution of the scintillator strips for a muon system at future colliders},
	journal = {Nuclear Instruments and Methods in Physics Research Section A},
	volume = {823},
	pages = {120--125},
	year = {2016},
	doi = {10.1016/j.nima.2016.03.091},
}

@article{Denisov2017,
	author = {Denisov, Dmitri and Evdokimov, Valery and Luki{\'c}, Strahinja and Uji{\'c}, Predrag},
	title = {Test beam studies of the light yield, time and coordinate resolutions of scintillator strips with {WLS} fibers and {SiPM} readout},
	journal = {Nuclear Instruments and Methods in Physics Research Section A},
	volume = {848},
	pages = {54--59},
	year = {2017},
	doi = {10.1016/j.nima.2016.12.043},
}

@article{Artikov2016,
	author = {Artikov, A. and Baranov, V. and Budagov, Yu. and Chokheli, D. and Davydov, Yu. and Glagolev, V. and Kharzheev, Yu. and Kolomoetz, V. and Shalugin, A. and Simonenko, A. and Tereshchenko, V.},
	title = {Optimization of light yield by injection of the optical filler into the co-extruded hole of plastic scintillation bar with {WLS} fiber in it},
	journal = {Journal of Instrumentation},
	volume = {11},
	pages = {T05003},
	year = {2016},
}

@article{Luo2025,
	author = {Luo, Guang and Yin, Xiao-Hao and An, Feng-Peng and Wang, Zhi-Min and Hor, Y. K. and Lu, Pei-Zhi and Li, Ru-Hui and Li, Yi-Chen and He, Wei and Wang, Wei and Xiao, Xiang},
	title = {Performance of plastic scintillator modules for top veto tracker at {Taishan Antineutrino Observatory}},
	journal = {Nuclear Science and Techniques},
	volume = {36},
	pages = {91},
	year = {2025},
	doi = {10.1007/s41365-025-01696-2},
}

@article{Bloise2023,
	author = {Bloise, C. and Ceravolo, S. and Cervelli, F. and Colao, F. and Cordelli, M. and Corradi, G. and Di Falco, S. and Diociaiuti, E. and Donati, S. and Ferrari, C. and Gargiulo, R. and Gioiosa, A. and Giovannella, S. and Giusti, V. and Hampai, D. and Happacher, F. and Martini, M. and Miscetti, S. and Morescalchi, L. and Paesani, D. and others},
	title = {Design, assembly and operation of a Cosmic Ray Tagger based on scintillators and {SiPMs}},
	journal = {Nuclear Instruments and Methods in Physics Research Section A},
	volume = {1045},
	pages = {167538},
	year = {2023},
	doi = {10.1016/j.nima.2022.167538},
}

@article{Artikov2024,
	author = {Artikov, A. and Baranov, V. and Boikov, A. and Chokheli, D. and Davydov, Yu. I. and Glagolev, V. and Simonenko, A. and Tsamalaidze, Z. and Vasilyev, I. and Zimin, I.},
	title = {High efficiency muon registration system based on scintillator strips},
	journal = {Nuclear Instruments and Methods in Physics Research Section A},
	volume = {1064},
	pages = {169436},
	year = {2024},
	doi = {10.1016/j.nima.2024.169436},
}

@article{britvich_radiation_1993,
	title = {Radiation damage studies on polystyrene-based scintillators},
	volume = {326},
	issn = {0168-9002},
	url = {https://www.sciencedirect.com/science/article/pii/016890029390849D},
	doi = {10.1016/0168-9002(93)90849-D},
	number = {3},
	urldate = {2026-10-02},
	journal = {Nuclear Instruments and Methods in Physics Research Section A: Accelerators, Spectrometers, Detectors and Associated Equipment},
	author = {Britvich, G. I. and Peresypkin, A. I. and Rykalin, V. I. and Vasil'chenko, V. G. and Kornylovskaya, L. D. and Malinovskaya, S. A. and Skripkina, V. T. and Shershukov, V. M. and Yushko, E. G. and Kulichenko, A. V. and Pyshchev, A. I.},
	month = mar,
	year = {1993},
	pages = {483--488},
}






\end{document}